\documentclass[conference,compsoc]{IEEEtran}
\IEEEoverridecommandlockouts
\usepackage{stfloats}
\usepackage{graphicx}

\usepackage{soul}

\usepackage{xcolor}

\usepackage{enumitem}

\usepackage{algorithm}
\usepackage{algorithmic}
\usepackage{booktabs, multirow}
\usepackage{threeparttable}

\usepackage{amsmath}

\DeclareMathOperator*{\argmin}{arg\,min}

\usepackage{url}
\usepackage{amssymb}
\usepackage{float}

\ifCLASSINFOpdf
\else
\fi
\begin{document}
%

\title{ALIF: Low-Cost Adversarial Audio Attacks on Black-Box Speech Platforms using Linguistic Features}



\author{
\IEEEauthorblockA {Peng Cheng\textsuperscript{$*, \#$}, Yuwei Wang\textsuperscript{$*, \#$}, Peng Huang\textsuperscript{$*, \#$}, Zhongjie Ba\textsuperscript{$*, \#, \ddagger$} \thanks{$\ddagger$ Corresponding Author: Zhongjie Ba}, Xiaodong Lin\textsuperscript{$\dagger$}, Feng Lin\textsuperscript{$*, \#$}, Li Lu\textsuperscript{$*, \#$}, 
Kui Ren\textsuperscript{$*, \#$}}
\IEEEauthorblockA{
\textsuperscript{$*$}Zhejiang University, Hangzhou, China\\
\textsuperscript{$\#$}ZJU-Hangzhou Global Scientific and Technological Innovation Center, Hangzhou, China\\ \textsuperscript{$\dagger$}University of Guelph, Guelph, Canada\\
\{peng\_cheng, yuwei.wang, penghuang, zhongjieba, flin, li.lu, kuiren\}@zju.edu.cn, xlin08@uoguelph.ca}
}

\markboth{Journal of \LaTeX\ Class Files,~Vol.~14, No.~8, August~2015}%
{Shell \MakeLowercase{\textit{et al.}}: Bare Demo of IEEEtran.cls for IEEE Transactions on Magnetics Journals}
%



\IEEEtitleabstractindextext{%
\begin{abstract}
Extensive research has revealed that adversarial examples (AE) pose a significant threat to voice-controllable smart devices. Recent studies have proposed black-box adversarial attacks that require only the final transcription from an automatic speech recognition (ASR) system. However, these attacks typically involve many queries to the ASR, resulting in substantial costs. Moreover, AE-based adversarial audio samples are susceptible to ASR updates. In this paper, we identify the root cause of these limitations, namely the inability to construct AE attack samples directly around the decision boundary of deep learning (DL) models. Building on this observation, we propose ALIF, the first black-box adversarial linguistic feature-based attack pipeline. We leverage the reciprocal process of text-to-speech (TTS) and ASR models to generate perturbations in the linguistic embedding space where the decision boundary resides. Based on the ALIF pipeline, we present the ALIF-OTL and ALIF-OTA schemes for launching attacks in both the digital domain and the physical playback environment on four commercial ASRs and voice assistants. Extensive evaluations demonstrate that ALIF-OTL and -OTA significantly improve query efficiency by 97.7\% and 73.3\%, respectively, while achieving competitive performance compared to existing methods. Notably, ALIF-OTL can generate an attack sample with only one query. Furthermore, our test-of-time experiment validates the robustness of our approach against ASR updates.
\end{abstract}


}

\maketitle

\IEEEdisplaynontitleabstractindextext

%
\IEEEpeerreviewmaketitle

\section{Introduction}\label{sec:intro}

Smart devices pervasively integrate voice control functionality. Users are getting used to interacting with smart devices with their voice to enjoy hands-free convenience. As a result, smart devices, such as smartphones, smart speakers, and automobiles, have adopted the voice assistant (VA) function to turn themselves into voice-controllable devices. In point of fact, over 132 million people used VAs in the US in 2021, and the number continues to grow~\cite{usersReport}. More than 4.25 billion VAs have been installed globally, projected to reach 8.4 billion by 2024~\cite{numberReport}.


The prevalence of voice-controllable devices brings security risks. Smart appliances take voice commands as input for performing heterogeneous actions, including security and safety-critical tasks, such as thermal adjustment, online payment, and even autonomous driving~\cite{Levinson2011Towards,Geiger2012Are}. Attackers can exploit the voice interaction to inject malicious speech commands without raising users' suspicions and cause severe security and safety consequences, including economic losses, privacy violations, health issues (e.g., thermometer overheating), bodily harm (e.g., car accidents), etc.

Injecting malicious commands covertly into voice-controllable devices has been realized with adversarial audio techniques~\cite{zhang2017dolphinattack, carlini2018AEs, Abdullah2019Practical, chen2020devilswhisper, zheng2021black}. The requirements of an attack audio are twofold: to maintain the covertness of the attack audio, it cannot sound like the intended command; to guarantee the attack's success, it should be correctly recognized as the intended command by a speech recognition model. The adversarial example (AE), which adds minute perturbation to original audios, naturally fits the attack requirements as the method does not affect users significantly and can affect the target model.

The problem of generating AEs to attack commercial voice-controllable devices is formulated as a black-box adversarial attack problem. The speech recognition models applied by commercial products are unknown to the attacker. To generate qualifying AEs, the adversaries usually query the black-box model thousands of times with trial audio examples, get the query results (i.e., transcription), and iteratively optimize the attack examples leveraging the transcriptions as optimization guide~\cite{Khare2019Adversarial,zheng2021black,Liu2022WhenEvilCalls}. 

The primary goal of black-box studies is to reduce attack cost, but the state-of-the-art solutions are still lacking in their suitability for practical applications. Considering the generation of each attack sample requires abundant queries and each one has a tangible monetary cost\footnote{300000 queries to the Google Cloud Speech-to-Text service (STT) result in a cost of about 1200 USD~\cite{Liu2022WhenEvilCalls}.}, existing works seek to improve the query efficiency and therefore reduce the attack price. Improving the query efficiency can also benefit the covertness of the attack. High-frequency querying with similar audio content may trigger the network defense system such as intrusion detection~\cite{Liu2022WhenEvilCalls,chen2020HopSkipJumpAttack,suya2020hybrid}. Recent works achieved 1500 queries on a commercial speech recognition model to generate one attack sample~\cite{chen2020devilswhisper,Liu2022WhenEvilCalls}, and the current level remains considerably distant from the efficient and practical standard, indicating a critical need for further advancement.

\begin{figure}[t]
    \centering
    \includegraphics[width=\linewidth]{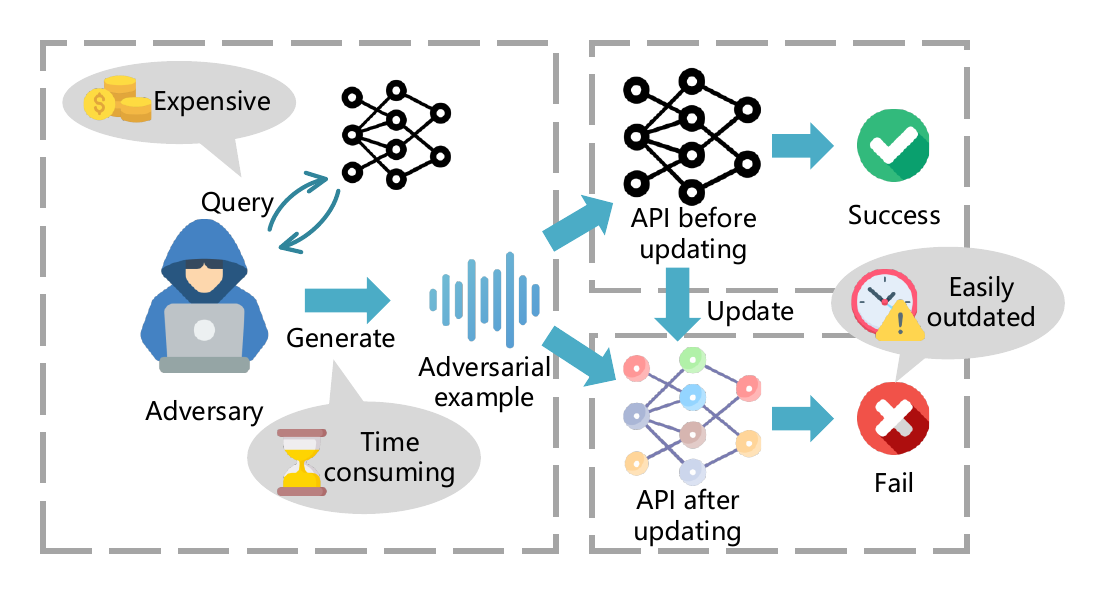}
    \caption{Limitations of existing adversarial attacks based on adversarial examples (AEs). The traditional pipeline for generating black-box audio adversarial examples necessitates the attacker making many API queries, which is both costly and time-consuming. The attacker can then obtain an example capable of successfully attacking the API. However, the example's effectiveness is significantly diminished by model updates.}
    \label{fig:introduction_problem_description}
\end{figure}

In addition to the low efficiency of current state-of-the-art techniques, we have observed that existing attacks based on AEs are susceptible to losing their efficacy when encountering model updates, resulting in highly exacerbated attack costs. Specifically, we found that AEs are sensitive to changes in the deep learning model, such as the automatic speech recognition model (ASR), and fine-tuning the model with new data may cause the previously trained AEs to lose their effectiveness, as demonstrated in Figure~\ref{fig:introduction_problem_description}. This sensitivity represents a common shortcoming of AE-based attacks. For instance, Devil's Whisper~\cite{chen2020devilswhisper} discovered that previously workable AE samples no longer worked with later versions of the model. As a result, manufacturers' unpredictable model optimization behavior will make AE-based attacks easily outdated, increasing the chances of attack failure. In case of failure, regenerating adversarial audio from scratch incurs substantial time and financial overhead, doubling or more the cost of the attack.

Reducing the cost of black-box adversarial attacks for command injection encounters significant challenges. Firstly, the limited knowledge of the ASR system. Existing works utilize advanced optimization methods, such as the Revolution Algorithm and Gradient Estimation Algorithm, to reduce query numbers. However, the lack of an explainable guideline and reliance on trial and error remain issues. Approximately 1500 queries are still required to find a satisfactory attack sample in a black-box setting. Secondly, the susceptibility of AEs to model updates lacks sufficient research and solutions. We identify that model fine-tuning changes decision boundaries, rendering previously trained adversarial examples ineffective.



In this paper, we propose a novel low-cost attack on black-box ASR using \underline{a}dversarial \underline{li}nguistic \underline{f}eatures (ALIF)\footnote{The demos are presented in https://taser2023.github.io/. The source code is available in https://github.com/TASER2023/TASER.}. The core idea is to generate adversarial audio samples from the decision boundary space of an ASR. Our approach is inspired by two observations: (i) current AE-based methods for adding perturbations result in query inefficiency and susceptibility to model updates. Existing adversarial attacks on ASRs involve adding perturbations to raw inputs (audio waveforms) to search for AEs around the ASR decision boundary. However, mapping raw inputs to lower-dimensional embeddings in the ASR model reduces perturbation efficiency. It fails to guarantee sufficient distance between the embedding and the decision boundary, resulting in susceptibility to model changes. (ii) This limitation can be addressed by constructing adversarial audio from the representation space of the decision boundary. We propose a novel approach that generates adversarial audio samples from a space similar to the one where the ASR decision boundary lies. ASR and text-to-speech (TTS) synthesis are two reciprocal processes. Specifically, the low-dimensional linguistic embeddings extracted from the text analysis module of TTS capture the semantics of the raw input (i.e., text) and largely overlap with the space of the ASR decision boundary. Perturbations on this TTS linguistic feature space can affect ASR recognition. Based on this insight, we generate adversarial perturbations on the linguistic features of a TTS model to construct the adversarial attack audio samples. Compared to existing attacks, our proposed ALIF pipeline significantly improves the query efficiency by approximately an order of magnitude while demonstrating resilience to model updates, resulting in significantly reduced attack costs.

ALIF can generate audio that is incomprehensible to humans yet interpretable by ASR platforms. The capability opens up various attack scenarios, including hidden command injection in the digital domain, such as deceiving online subtitling services, and physical attacks in the real world to control VAs for unexpected command execution. Specifically, we introduce two attack schemes based on ALIF: ALIF over-the-line (ALIF-OTL) for the adversarial attack in the digital domain and ALIF over-the-air (ALIF-OTA) for the attack in the physical environment.

\noindent \textbf{Contributions.} Our contributions are as follows.
\begin{itemize}[leftmargin=*, topsep=0pt]
\item To the best of our knowledge, this paper is the first to study the generation of black-box adversarial audio from the linguistic feature space of TTS.
\item We propose ALIF, a black-box attack pipeline based on adversarial linguistic features against commercial ASR platforms. Depending on the pipeline, ALIF-OTL and ALIF-OTA are proposed for the digital and over-the-air attack scenarios. A new platform attack scenario that fools the online subtitling service is also proposed.


\item We conduct comprehensive experiments against commercial ASRs and well-known VA products to validate the effectiveness and practicality of our adversarial attacks. ALIF-OTL can generate adversarial audio with an average of 35 queries for the attack in the digital domain (only one query is also feasible), achieving 97.7\% query efficiency improvement over existing works and at least an average success rate of 95.8\% on 4 ASR APIs. ALIF-OTA achieves 73.3\% query efficiency improvement on state-of-the-art works and obtains an 81.3\% success rate on the APIs. Experimental results show our adversarial audio samples are robust against attack environments, attack distance, and model updates. Evaluations of various impact factors verify the practicality of our ALIF attacks.



\end{itemize}

\section{Background}~\label{sec:background}
This section introduces speech recognition/synthesis and the conventional black-box AE attack method.
\subsection{Automatic Speech Recognition}
\begin{figure}[tbp]
    \centering
\includegraphics[width=\linewidth]{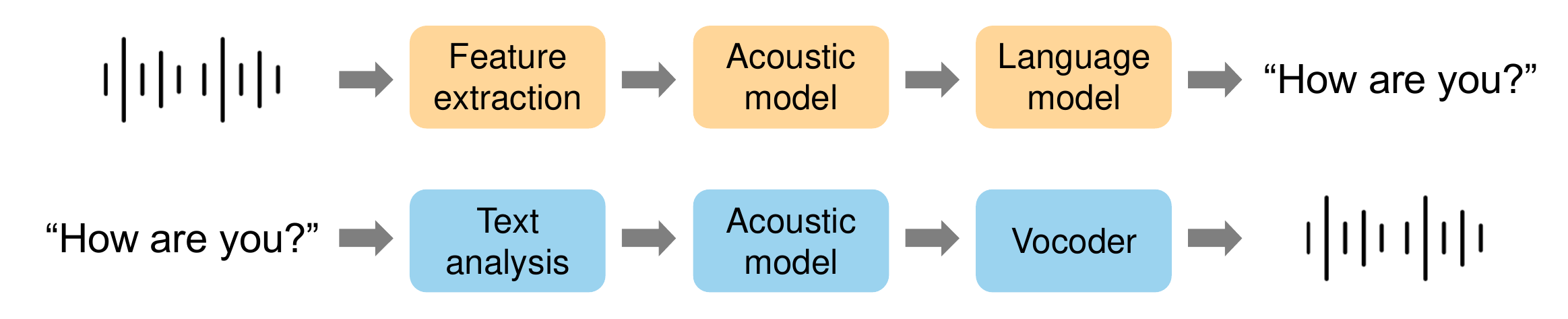}
    \caption{Architectures of an ASR and a TTS system.}
    \label{fig:ASRandTTS}
\end{figure}

An ASR system takes human speech signals as the input and produces the semantic meaning of the spoken content in the form of text. We denote the input waveform as $x$, the ASR system as $f(\cdot)$, and the output transcription as $y$. The function of an ASR is formulated as $y = f(x)$. The upper part of Figure~\ref{fig:ASRandTTS} shows a typical ASR system's architecture consisting of feature extraction, acoustic model, and language model. Classical ASR applies a Gaussian Mixture Model-Hidden Markov Model (GMM-HMM) to build the acoustic model. With the rapid development of deep learning, modern ASRs universally apply neural networks as their core and achieve outstanding recognition performance. Depending on the model architecture and neural network variants (e.g., CNN and RNN) used, a wealth of ASR categories exist. Introducing different ASR models in detail is out of the scope of this paper, and we present the representative components instead. The feature extraction component extracts acoustic features essential to the semantic information; then, the acoustic model predicts the most probable linguistic units based on the features. Finally, the language model generates the text sequence with the highest probability. 





\subsection{Speech Synthesis}
Speech synthesis, also known as Text-To-Speech (TTS), aims to synthesize natural and intelligible speech signals given an arbitrary sample of text as input~\cite{tan2021survey}. The task can be formulated as $x = f(y)$. The lower part of Figure~\ref{fig:ASRandTTS} shows the architecture of a typical TTS model consisting of three components: text analysis, acoustic model, and the vocoder. At the text analysis stage, the system encodes the text to the linguistic feature embedding space. Then the embedding is fed to the acoustic model to obtain the acoustic features (usually a Mel spectrogram). Finally, the vocoder converts the spectrogram to an audio waveform.

\subsection{Black-Box Adversarial Audio Attacks}
\noindent\textbf{Basic Adversarial Examples.} The goal of an AE-based attack is to deceive the ASR system into transcribing input waveform $x$ incorrectly by adding perturbations $\delta$, resulting in a transcription that does not match the correct transcription $y$. In a non-targeted attack, the requirement is that $SR(x+\delta) \neq y$, where SR denotes the recognition function of the ASR. In contrast, in a targeted attack, the condition is that $SR(x+\delta) = T$, where $T$ represents the malicious command the adversary intends to activate. Since AEs typically require the perturbations to be imperceptible, the amplitude of $\delta$ is usually constrained. The formulation of AE-based attacks can be expressed as:


\begin{equation}\label{eq:whiteboxAE}
    \argmin_{\delta}{L(SR(x+\delta), T) + \alpha \cdot dB_x\delta}
\end{equation}
\noindent where $L$ denotes the loss function indicating the similarity between the attack audio transcription and the target text. $\alpha$ is the weight to trade off being covert and adversarial. To solve the problem, a loss value from $L$ is calculated, and gradient descent is applied to find the optimal $\delta$. The gradient calculation requires information about the ASR architecture and parameters, which is impractical in the real world, where the specifics of commercial ASR are unknown.

\noindent \textbf{Black-Box Adversarial Examples.} In a black-box setting, internal knowledge, such as the weights of the target ASR, is unknown. The attacker can only obtain the final transcription, therefore, cannot calculate $L$, nor can they
apply gradient descent to solve for the optimal $\delta$. To ensure the intended command can be activated, the AE training process usually starts from the target command audio sample and gradually modifies the sample towards another unsuspicious signal $x$, aiming to generate an attack audio sample that sounds benign but is recognized as the target command. The problem in Formula~\ref{eq:whiteboxAE} is reformed as
\begin{equation}\label{eq:blackboxAE}
    \argmin_{\delta}L = 
    \begin{cases}
    \|\delta\|_p, &\textbf{$SR(x+\delta) = T$}\\
    +\infty, &otherwise
    \end{cases}
\end{equation}
\noindent where $\|\delta\|_p$ is a norm function to limit the perturbation, and $L$ is the objective function. Given the opaque ASR model, the optimization is based on heuristic techniques, which make no assumption about the model, search a large optimization space, and develop iteratively with experience learned from the previous round. The experience is learned from the query results. Such an iterative process normally requires extensive queries. Related studies have utilized evolution algorithms~\cite{Taori2018Adversarial, zheng2021black} and gradient estimation methods~\cite{Liu2022WhenEvilCalls} as the training methodology.

\section{System and Threat Models}

We demonstrate our attack schemes for both the digital domain and the physical environment, namely over-the-line (OTL) and over-the-air (OTA) scenarios. Leveraging the ALIF pipeline, we employ different training methods to generate adversarial audio samples against commercial ASRs in each scenario. The method for generating the adversarial attack in the OTL scenario is called ALIF-OTL. Notably, we propose a new type of digital attack called the \emph{platform attack} in the OTL scenario, where the online subtitle transcribing service is misled. The method for generating the adversarial attack in the OTA scenario is called ALIF-OTA. This section presents the system and threat models of ALIF-OTL and ALIF-OTA. 

\begin{figure}[!t]
\centering
\includegraphics[width=0.8\columnwidth]{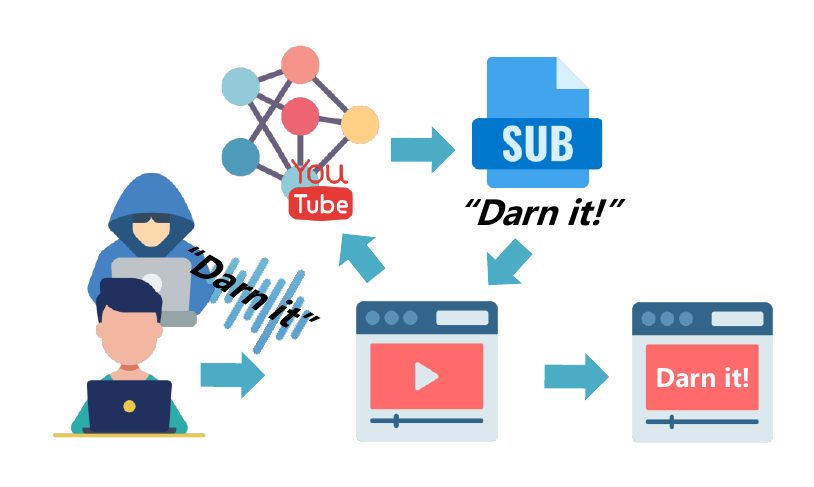}
\caption{\label{fig:systemModel} System model of ALIF-OTL. This depicts an attacker incorporating an adversarial audio track into a video. As a result, when the manipulated video is uploaded to the platform, the automatic subtitling service inadvertently generates inappropriate text. The ALIF-OTL attack occurs within the digital domain.}

\end{figure}


\subsection{System and Threat Model of ALIF-OTL}
\noindent\textbf{System Model.} The system model of ALIF-OTL is presented in Figure~\ref{fig:systemModel}. Online video and multimedia content platforms (e.g., YouTube) have started providing a subtitling service, which supports an automatic transcribing function for on-demand and live media content. These platforms apply an ASR to recognize the semantics of the audio track and generate subtitles. Audiences can opt-in to the service through a corresponding option in the playback agent, which allows subtitles to be shown as the video plays. As Amazon introduces~\cite{AmazonTranscribe}, subtitling content helps improve accessibility and engagement. It can also be a compliance requirement for video programming distributors to support hard-of-hearing users. In addition to utilizing the video platform's own service, content creators can use third-party services (e.g., Amazon Transcribe) to add subtitles for their media content.

\noindent\textbf{Attacker's Goal.} 
The attacker desires to launch an adversarial audio \emph{platform attack}. The goal is to damage the reputation of media content providers or cause trauma to audiences of particular communities. To achieve this goal, the attacker delivers the adversarial audio as the audio track of a video. The attacker can deceive the ASR of either a third-party subtitling service or the native feature of a video content platform like YouTube, leading to the generation of biased transcription text as the video subtitle. Users will see disinformation created intentionally by the attacker when the subtitle is shown. Since users only hear stuttering sounds and cannot understand the content, they would intuitively think it is caused by \emph{network jitter/packet loss}. However, these inappropriate transcriptions can inflict trauma on viewers and cause them to have negative experiences with the platform, possibly damaging the platform's reputation. 

Considering the expense and time overhead of querying commercial online APIs, the attacker would like to keep the cost as little as possible. Meanwhile, he/she would like to ensure the performance stability of the generated attack audio samples, namely reducing the risk that the audio becomes ineffective due to unpredictable ASR model updates by the service provider. A typical failure occurs when: the attacker obtains adversarial audio signals after many queries and time-consuming training but finds them ineffective during the attack.

\noindent \textbf{Attacker's Capability} The attacker is restricted from launching the attack in the black-box setting. ASRs typically do not provide the confidence score for transcriptions because it does not benefit the user's experience~\cite{zheng2021black}. The attacker also does not know the training dataset of the ASR, and he/she must generate the adversarial audio using a TTS model. The attacker can query the target ASR (e.g., Amazon, Microsoft, iFLYTEK, etc.) and prepare attack audio accordingly.

\begin{figure}[!t]
\centering
\includegraphics[width=1\columnwidth]{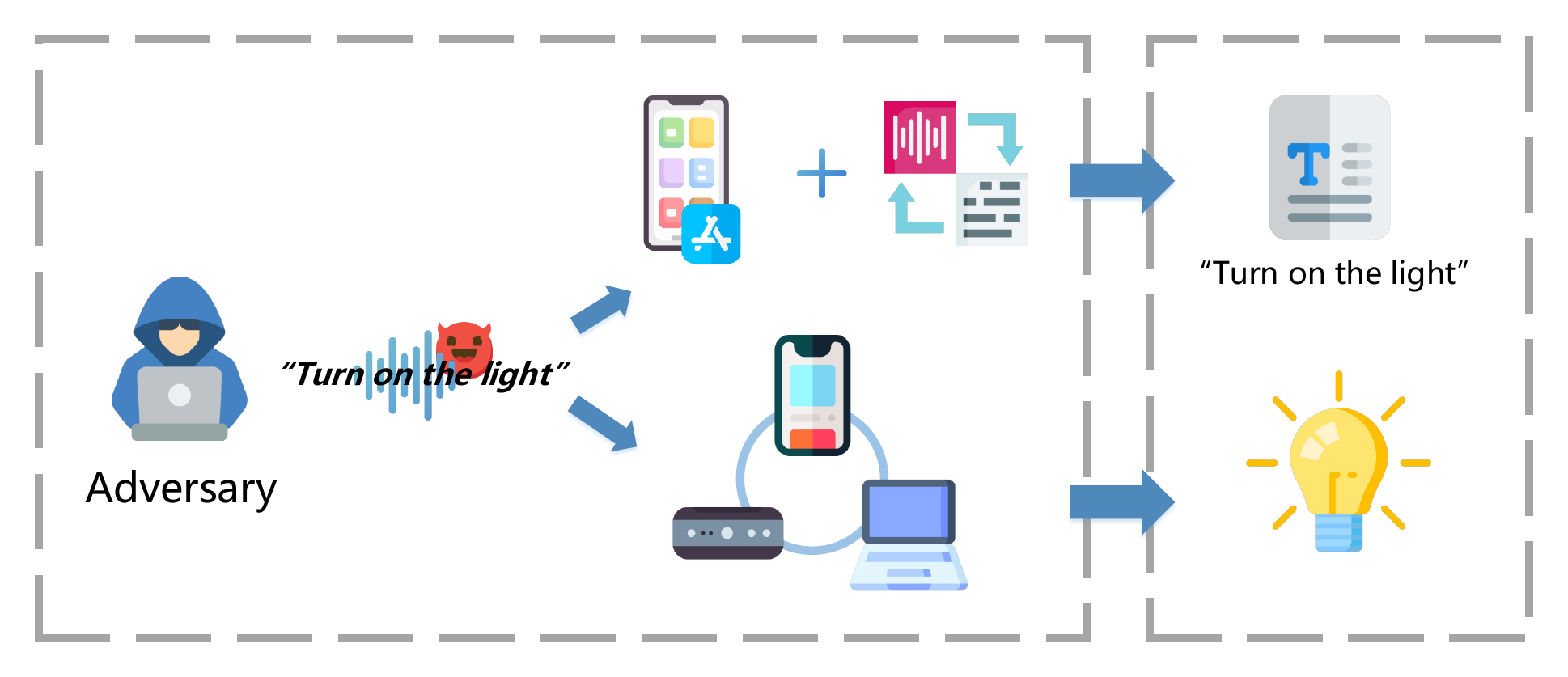}
\caption{\label{fig:OTAsystemModel} System model of ALIF-OTA. This model illustrates two potential attack scenarios. In the first scenario (upper part), the smartphone apps with voice interaction capabilities record the attack audio signals in the environment and then call online APIs to transcribe audio. In the second scenario (lower part), an attacker plays attack audio samples to activate voice assistants - such as smart speakers - that possess their own ASR backend, thereby executing the commands.}
\end{figure}

\subsection{System and Threat Model of ALIF-OTA}
\noindent\textbf{System Model.} Figure~\ref{fig:OTAsystemModel} presents the system model of ALIF-OTA, which represents a typical adversarial audio attack scenario. In the OTA scenario, the attack targets can be categorized into two types: commercial online ASR APIs and VAs (such as smart speakers and virtual assistants on computers). Our study specifically includes online ASR APIs to account for practical considerations, as small-scale companies may purchase online APIs rather than develop their own ASR system.

\noindent\textbf{Attacker's Goal.} The attacker intends to attack the VA running on a voice-controlled smart device, causing the ASR to transcribe the hidden command from the incomprehensible audio, triggering the device to execute unintended actions. 

\noindent \textbf{Attacker's Capability.} We assume the adversary can launch the attack in the physical environment when the owner is not using the VA device, similar to Metamorph~\cite{Chen2020Metamorph}. Even if the owner is in proximity and hears the attack audio, he/she cannot recognize the sound as a speech command and would not realize the onset of an adversarial attack, as described in Carlini et~al.~\cite{Carlini2016Hidden} and Abdullah et~al.~\cite{Abdullah2019Practical}. The attacker can launch the attack through a covertly placed speaker or a compromised speaker, following the convention in Devil's Whisper~\cite{chen2020devilswhisper} and OCCAM~\cite{zheng2021black}.

\section{Motivation and Our Design}\label{sec:motivation}
Existing black-box attacks suffer from query inefficiency (except for NI-OCCAM~\cite{zheng2021black}, a non-interactive attack method). Additionally, they exhibit susceptibility to model updates, as outlined in Section~\ref{sec:intro}. These inefficiencies lead to substantial attack costs. Next, we introduce our proposed design to address these issues.

The vulnerability of deep learning (DL) models to imperceptible perturbations has attracted much attention since the works of Szegedy~et~al.~\cite{szegedy2013intriguing} and Biggio et~al.~\cite{Biggio2013Evasion}. A recent study by Shamir et~al.~\cite{shamir2021dimpled} proposes a dimpled manifold model (DMM), which provides a better explanation for the working mechanism of AE. The effect of AE is closely related to the model's decision boundary. During model training, the initially randomly oriented decision boundary quickly aligns to a low-dimensional manifold that contains the representation embedding of all training samples. In the second training phase, the decision boundary starts dimpling, and \emph{shallow} bulges are generated to move the decision boundary towards the right direction around the data embedding according to the labels. When a service provider fine-tunes its ASR model with newly-collected training data, such as noisy data for robustness improvement, the decision boundary undergoes these two training phases again, resulting in the decision boundary reforming around the new sample embeddings. This reformulation of the decision boundary renders previous AEs invalid with a high probability. Existing AEs add perturbations in the raw input space rather than in the lower-dimensional representation space, which does not guarantee enough distance between the attack sample embedding and the decision boundary. As a result, AEs are easily affected by changes in the decision boundary induced by model updates. To address these challenges, we propose ALIF, a novel attack scheme that reduces inefficient querying and significantly improves reliability.

\section{ALIF-OTL: Over-the-Line ALIF Attack on ASR APIs}
In this section, we present a detailed description of the ALIF-OTL algorithm. It is important to note that ALIF serves as the foundational pipeline for both ALIF-OTL and ALIF-OTA, with the two attack schemes differing primarily in their training methodologies. We introduce the ALIF pipeline in the ALIF-OTL demonstration.

\begin{figure}[t]
\centering
\includegraphics[width=0.8\columnwidth]{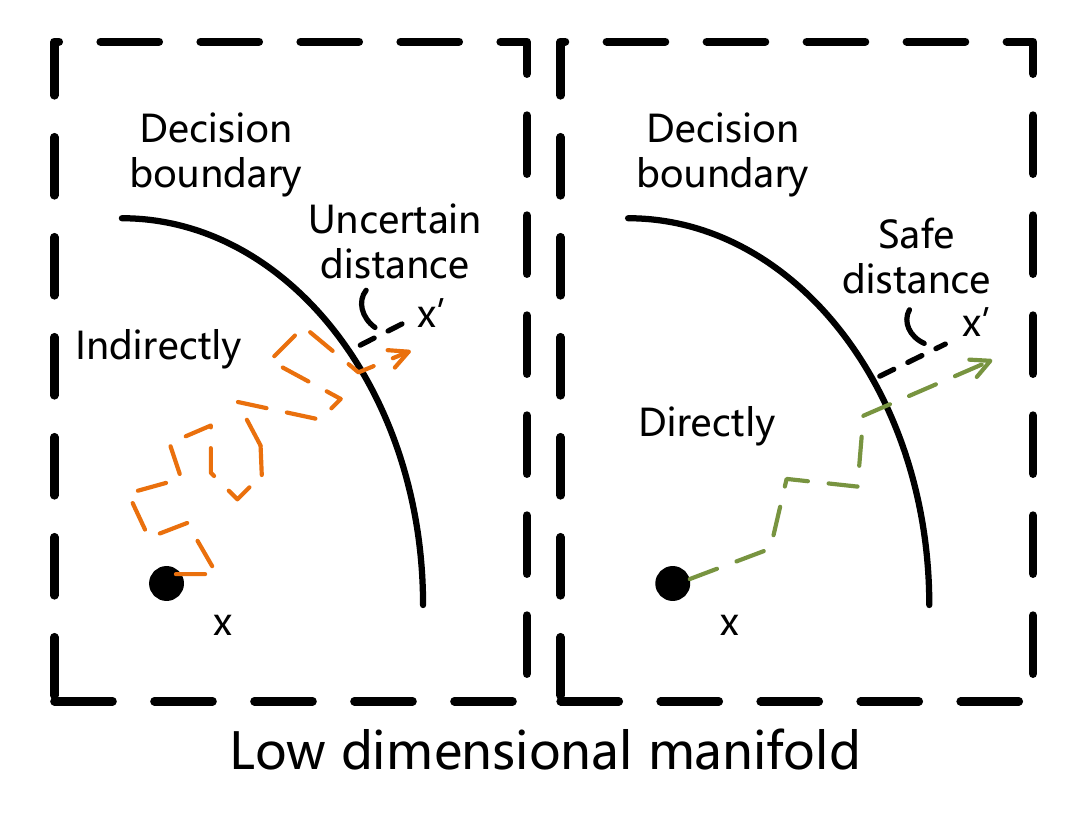}
\caption{\label{fig:contrastofAEandOurs} Contrast of AE-based attacks and our work.}
\end{figure}

\begin{figure*}[tbp]
    \centering
\includegraphics[width=0.9\linewidth]{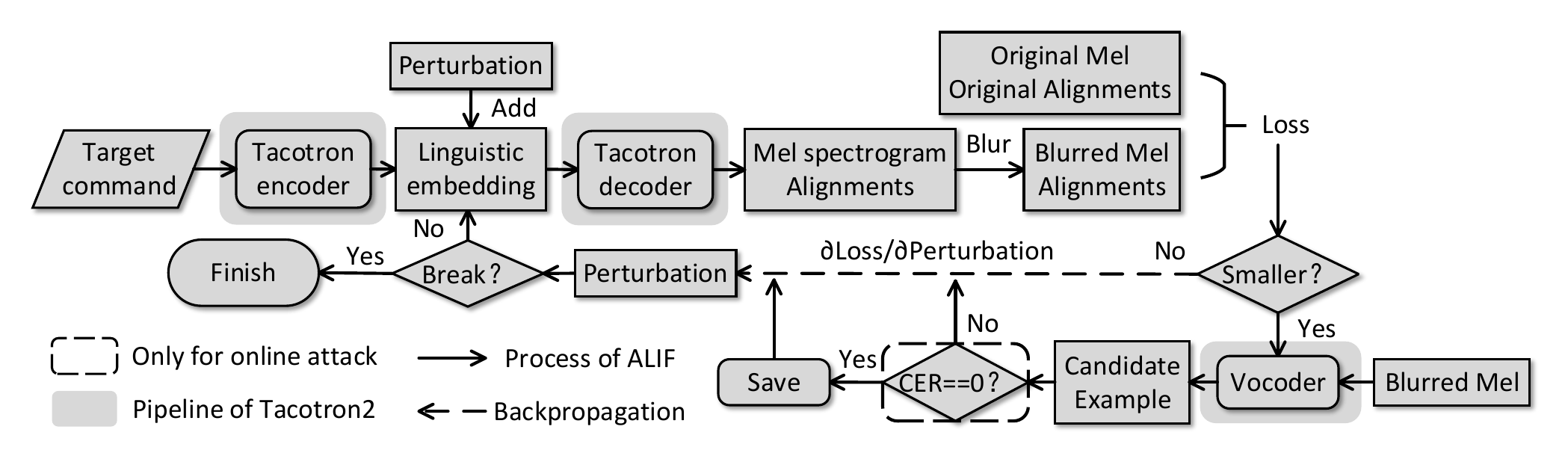}
    \caption{Architecture of the ALIF-OTL scheme. The scheme is based on the ALIF pipeline shown with a solid line.}
    \label{fig:alifSystem}
\end{figure*}

\subsection{Design Intuition}
As revealed and verified by a massive number of papers (e.g., Ruderman~et~al.~\cite{ruderman1994statistics}, Pope~et~al.~\cite{pope2021the}, and Shamir~et~al.~\cite{shamir2021dimpled}), real-world data distribution can be described by low dimensional structure (i.e., a manifold). The low-dimensional manifold is easier for neural networks to learn and form complex decision boundaries from a relatively small amount of training samples. According to DMM, the decision boundary evolves primarily based on the low dimensional representations of the input sample (i.e., images or text), which heavily affect the establishment and the counterintuitive properties of adversarial examples. As the left of Figure~\ref{fig:contrastofAEandOurs} shows, existing black-box AE training adds perturbation to the raw input to shift the low dimensional representation, which is an indirect optimization process. Because the process lacks accurate guidance from the gradient descent method, the indirect search method is inefficient. Regarding the vulnerability to model updates, existing works~\cite{Liu2022WhenEvilCalls, zheng2021black} iteratively optimize attack audio samples in the input space (see Section~\ref{sec:background}). The process stops when the pre-defined maximum number of queries is reached. Since the raw input space is not where the decision boundary lies, this process cannot ensure a large enough distance between the AE and the decision boundary, which increases the risk of attack audio failure when the decision boundary changes. To address the bottlenecks, our key idea is to construct perturbations directly from the decision boundary space and make the distance between attack samples and the boundary farthest allowed under certain constraints (i.e., right part of Figure~\ref{fig:contrastofAEandOurs}). 





Considering the architectures of ASR and TTS models in Figure~\ref{fig:ASRandTTS}, we can consider the two models to be reciprocal processes. Both ASR and TTS have a key component called the acoustic model. The purpose of this component in the two models is the same: to map the relationship between acoustic features and linguistic features. In TTS, the acoustic model uses linguistic feature as the input, which is the representation embedding of the linguistic feature space~\cite{GoogleMachineLearning}. The linguistic embedding captures the essential semantic information of the text. Because the transcription of ASR is also formed based on the linguistic units contained in the output of the acoustic model, the TTS linguistic feature space is closely correlated with its counterpart in the ASR. This correlation between TTS and ASR is the observation concluded from our experiment and existing works. Existing studies have widely adopted TTS to synthesize audio commands~\cite{chen2020devilswhisper, zheng2021black, Liu2022WhenEvilCalls}. Intuitively, we seek to generate ``adversarial examples" from the input space of the acoustic model and design adversarial linguistic feature-based attacks. The generated ``adversarial examples" in our scenario can be called ``adversarial embeddings".

\begin{figure*}[!t]
\centering
\includegraphics[width=0.9\linewidth]{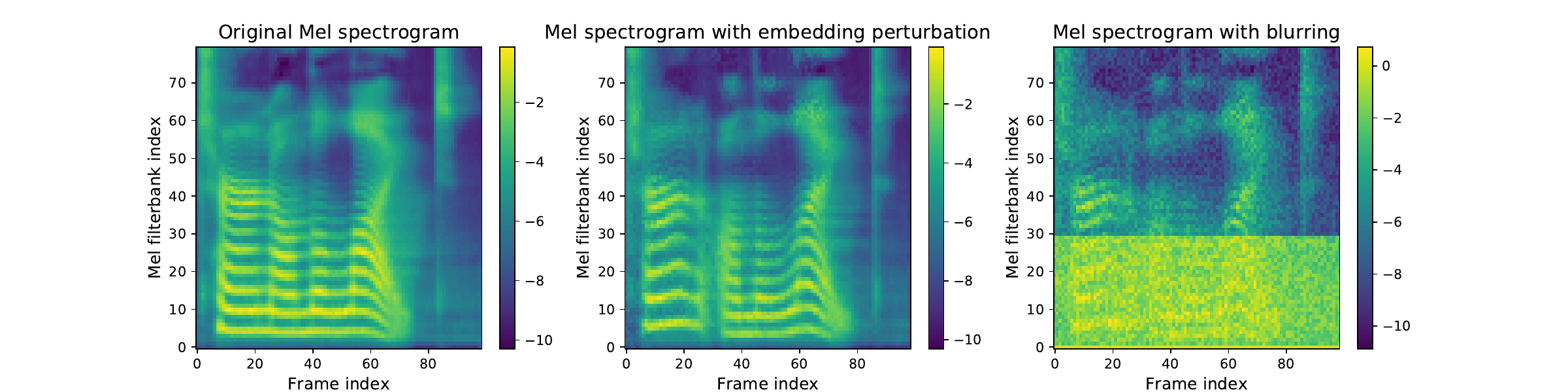}
\caption{\label{fig:contrastonPreandAfter} Comparison between the original spectrogram, after perturbing the embedding, and after the blurring operation.}
\end{figure*}


\subsection{Our Method}
\textbf{Design Overview.} ALIF-OTL creates incomprehensible audio to attack the subtitling APIs online. A conventional TTS generates a Mel spectrogram, which represents text content, then applies a vocoder to the spectrogram to synthesize signals representing the intended speech. We add perturbation to the representation of linguistic feature space to achieve a similar goal as hidden command attacks~\cite{Carlini2016Hidden, Abdullah2019Practical}, which is to make the attack audio not sound like the command anymore but still recognizable as a command. As a result, users will stay unaware of the occurrence of an attack.


Figure~\ref{fig:alifSystem} shows the ALIF-OTL attack scheme. Our backbone model for audio generation is based on a well-known TTS called Tacotron2~\cite{shen2018Tacotron}, shown in the shaded part of Figure~\ref{fig:alifSystem}. The output of the Tacotron2 decoder consists of the predicted Mel spectrograms and the alignments. The former is the input of the vocoder, while the latter is used to predict whether the output sequence has been completed. We use a similar loss function to Tacotron2, which includes $L_{Mel}$ and $L_{Gate}$. $L_{Mel}$ is used to measure the difference between the original Mel spectrogram and the adversarial one, while $L_{Gate}$ controls the adversarial audio to have a similar length to the benign one. We will introduce them later.

To make an audio sample highly-distorted that individuals cannot understand, we optimize the perturbations using gradient descent to minimize the loss function. We observe that the magnitude of the perturbation is negatively correlated with the text similarity. The higher the amplitude of the perturbation, the more significant the difference between the transcription of the attack audio and the target content, which aligns with intuition. Based on this observation, we propose two variants of attacks: online and offline. We generate ten candidate adversarial embeddings for a target command and optimize each embedding vector for 50 iterations. Within this process, we query the target ASR API and calculate the character error rate (CER) to check whether the transcription is the same as the target command. During the \emph{online attack}, we only query the target API with audio samples if the loss calculated between the original and perturbed spectrograms is reduced. In contrast, we only query the API once at the last iteration round for \emph{offline attack}. If the amplitude of the perturbation exceeds the pre-defined threshold during the loop, the process of both attacks is terminated.







\noindent \textbf{Mel spectrogram blurring.} Prior to computing $L_{Mel}$, we execute a blurring operation on the audio spectrogram to further reduce perceptibility. The key idea is to disturb the spectral structure of the spectrogram while keeping the essential semantic-related part untouched. The output of the Tacotron decoder in Figure~\ref{fig:alifSystem} is the spectrogram that reflects the energy distribution in different frequencies. The following vocoder generates phase information based on the spectrogram and transforms the spectrogram into an audio waveform. We choose to manipulate the Mel spectrogram rather than the vocoder because the signal energy distribution in a spectrogram presents information about sound types. Vowels and consonants are the two major sound types present in speech, and their distinctiveness contributes to the intelligibility of a sentence~\cite{tahmina2014Perceptual,fogerty2012role}. To manipulate the energy distribution while ensuring transcription correctness, we empirically design three blurring techniques: 1) \emph{Decreasing the low-frequency energy.} We multiply the amplitude of frequencies for the first 30 Mel filter banks (80 filter banks in total) by a scalar factor $\alpha$ (e.g., 0.25 or 0.3). 2) \emph{eliminating very low frequencies.} The very low frequencies do not have a significant influence on the recognition results. We set the amplitude of the first $\beta$ Mel filter banks to zero. 3) \emph{superimpose uniform noise.} At last, we add a layer of noise to the Mel spectrogram, which effectively reduces the ``sharpness" of the audio. Specifically, we create a noise matrix with values sampling from a uniform distribution ranging from $-\gamma$ to $\gamma$ ($\gamma > 0 $) and with the same size as the attack audio, then overlap these two matrices. Figure~\ref{fig:contrastonPreandAfter} illustrates the spectrograms of audio before and after the blurring steps. 





\noindent \textbf{ALIF-OTL Formulation.} Our goal is to reduce the intelligibility of the audio while maintaining the correct transcription of the target content to activate the intended command. We calculate the difference between the original audio signal and the perturbed audio sample regarding both the Mel spectrogram (i.e., $L_{Mel}$) and the alignment vectors (i.e., $L_{Gate}$) . Specifically, we use the summed mean squared error (MSE) and BCEWithLogitsLoss~\cite{shen2018Tacotron} for measuring the difference, respectively. As a result, the objective functions are as follows: 
\begin{equation}\label{eq:TASERAE-OTL2}
    L_{Mel}=-MSE(Blur(D_{Mel}(Emb+\delta), hp), D_{Mel}(Emb)),
\end{equation}
\begin{equation}\label{eq:TASERAE-OTL3}
    L_{Gate}=BCE(Sigmoid((D_{Gate}(Emb+\delta)), D_{Gate}(Emb))
\end{equation}
\begin{equation}\label{eq:TASERAE-OTL}
    L_{Final} =  L_{Mel}+L_{Gate}
\end{equation}

\noindent where $L_{Mel}$ is the spectrogram difference. $D_{Mel}(\cdot)$ referes to the decoder function in Tacotron2 which uses the embedding to predict the corresponding spectrogram~\cite{shen2018Tacotron}. $D_{Gate}(\cdot)$ is vector representing alignment status. $Emb$ means the original representation of the target command in the latent space of linguistic features. $Emb+\delta$ is the adversarial embedding. $hp$ denotes the hyperparameters $\alpha$, $\beta$ and $\gamma$.

Details of ALIF-OTL is shown in Algorithm~\ref{alg:taser-otl}, $Thr$ is the pre-defined upper limit of the perturbation amplitude, which ensures the semantics are not shifted far away from the origin. The perturbations $\delta$ are represented as a matrix of values which are added to the embedding space. $\delta$ is measured by its infinity norm; that is
\begin{equation}
    \|\delta\|_p = \max{\delta_{ij}}
\end{equation}
\noindent where $\delta_{ij}$ is the element of the perturbation matrix $\delta$. $i$ denotes the index of the token (i.e., letter) in the text, and $j$ means the index of the element in the representation embedding of this token.




	
\begin{algorithm}
    \footnotesize
    \renewcommand{\algorithmicrequire}{\textbf{Input:}}
    \renewcommand{\algorithmicensure}{\textbf{Output:}}
	\caption{ ALIF-OTL }~\label{alg:taser-otl}
	\label{algUndersampling} 
	\begin{algorithmic}[1]
	
		\REQUIRE Target command $T$; target API $SR$; encoder $E$ and decoder $D$ of Tacotron2; vocoder V; threshold $Thr$; the blurring parameters $\alpha$, $\beta$, $\gamma$ and the number of iterations for one attack example $EpochMax$. 
		\ENSURE $AttackExample$
        \STATE $\delta = 0; BestLoss = +\infty; Epoch = 0$
        \STATE $Emb = E(T)$
        \STATE $OriSpec = D(Emb)$
        \STATE $OriginalAudio = V(OriSpec)$
        \STATE $AttackExample = OriginalAudio$
        \WHILE{$Epoch < EpochMax$}
            \STATE $AdvSpec = Blur(D(Emb + \delta), \alpha, \beta, \gamma)$
            \STATE Calculate $L_{Final}$ by Eq.\ref{eq:TASERAE-OTL2}, \ref{eq:TASERAE-OTL3} and \ref{eq:TASERAE-OTL}
            \STATE Update $\delta$ by gradient descent
            \IF{$L_{Final} < BestLoss$}
                \STATE $ BestLoss=L_{Final}$
                \STATE $CandidateExample = V(AdvSpec)$
                \STATE $t = SR(CandidateExample)$
                    \IF{$t == T$}
                        \STATE $AttackExample = CandidateExample$
                    \ENDIF
                    \IF{$\|\delta\|_p>Thr$}
                        \STATE break
                    \ENDIF
            \ENDIF
        \STATE $Epoch++$
        \ENDWHILE
        \RETURN $AttackExample$
	\end{algorithmic} 
\end{algorithm}


\section{ALIF-OTA: Over-the-Air ALIF Attack on Voice Assistants}

\subsection{Technical Challenges}
Because ALIF-OTL does not consider the impact factors of physical playback, it cannot be applied to in the OTA scenario. One big challenge is generating attack audio samples that can overcome the distortion brought by the physical playback environment. In traditional AE attacks, the attacker can incorporate OTA impact, such as reverberation, into the training objective function in the perturbation generation stage to achieve a ``rehearsal" effect. This processing significantly improves the attack robustness in the physical playback. However, for the ALIF attack, the origin of our optimization is the low-dimensional embedding, which is not the natural waveform domain. It is unknown how to reflect the reverberation and additional noise impact. Therefore, it is necessary to find an optimization solution that can simultaneously achieve intelligibility reduction and physical inference endurance.


\subsection{Our Method}
The key idea is to guide the optimization direction in a way that physical interference is considered. Meta-heuristic algorithms are suitable in this context. Inspired by Xie~et~al.~\cite{Xie2022universal}, we solve the problem with Particle Swarm Optimization (PSO). 


\noindent \textbf{Particle Swarm Optimization.} PSO is a kind of metaheuristic algorithm which is designed to solve nonlinear functions~\cite{eberhart1995new}. It requires no knowledge and does not need the problem to be differentiable. In more detail, the algorithm first initializes a population of random solutions. Every solution can be regarded as a particle $X_i=\{x_{i1}, x_{i2}, ..., x_{id}\}$ and the value of the solution is the position of the particle in the hyperspace. Meanwhile, each particle has a randomized velocity $V_i=\{v_{i1}, v_{i2}, ..., v_{id}\}$ and then they ``fly” iteratively through the hyperspace. Specifically, during iteration, the global best position and the personal best position of each particle are recorded. Then in each iteration, the velocity and position of each particle can be updated as follows:
\begin{equation}\label{eq:pso1}
v_{ij}=w\cdot v_{ij}+c_1\cdot r_1\cdot (pbest_{ij}-x_{ij})+c_2\cdot r_2\cdot(gbest_j-x_{ij})    
\end{equation}
\begin{equation}\label{eq:pso2}
x_{ij}=x_{ij}+v_{ij}
\end{equation}
where $v_{ij}$ means the $j^{th}$ dimension of the velocity of $i^{th}$ particle in the current iteration. $pbest$ and $gbest$ are the personal best position and global best position. $r_1$ and $r_2$ are two random uniformly distributed numbers between $0$ and $1$. $w$ is the inertia weight while $c_1$ and $c_2$ are two acceleration constants. In this work, we regard the noise added to the linguistic embedding as a particle. Unlike the original PSO, we initialize all the particles with a zero value to help them find the initial best position. We randomize $r_1$ and $r_2$ at every iteration for every particle. In order to reduce the computing overhead, each dimension of the same particle shares these two random numbers. We also set a threshold and only update the particle when it does not exceed the threshold. As a result, Eq.~\ref{eq:pso2} becomes:
\begin{equation}\label{eq:pso3}
x_{ij}=x_{ij}+v_{ij}, \text{ when }max(x_{ij}+v_{ij})<threshold
\end{equation}



\noindent \textbf{Design Overview.} 
Our goal in the OTA case is more challenging than the OTL case becuase of additional interferences in the physical playback environment.  Based on the same ALIF pipeline shown in Figure~\ref{fig:alifSystem}, we use PSO to search for the optimal perturbation iteratively. The perturbation vector $Emb$ is the particle, and the set of the embedding range consists of the optimization space. The ALIF-OTA scheme differs from the OTL counterpart in an extra noise overlay step. We intentionally create and combine the white noise with the generated waveform. The particles recognized correctly under this constraint have a better chance of succeeding in the OTA scenario. We do not consider the reverberation influence as we empirically find the effect is not severe in a residential room environment. Having obtained the attack audio waveform, we generate and add a white noise segment with the maximum magnitude of $\eta$ (e.g., 0.05 or 0.1) to the normalized audio $x$. The maximum value of the mixed audio becomes $1+\eta$ (1.05/1.1). We then normalize each waveform data point and multiply it by 10000. The processed audio is fed into an online ASR API to obtain the recognition result (i.e., $y$). The above parameter setting is empirically concluded with experiments. 




\noindent \textbf{ALIF-OTA Formulation.} To begin, we initialize 20 particles (perturbations) as zero matrices. Each particle then performs ten iterations to search for its best position. In each iteration, we query the API with and without the noise addition and record the loss $L_{Final}$ of those particles, which results in the correct transcription. Upon reaching the maximum number of iterations, we pick the particle with the minimum loss as the perturbation for the attack. In short, the ALIF-OTA problem is formulated as 
\begin{equation}\label{eq:TASERAE-OTA}
\begin{split}
    L_{Final} &=
    \begin{cases}
    L_{Mel}+L_{Gate}, & \text{if } condition1 \text{ and } condition2\\
    +\infty, & otherwise
    \end{cases}\\
    & while \; \;\;\|\delta\|_p \leq Thr, \\
    & AdvSpec=Blur(D(Emb+\delta), \alpha, \beta, \gamma), \\
    & condition1: \textbf{$SR(V(AdvSpec)) = T$}, \\ 
    & condition2: \textbf{$SR(V(AdvSpec)))+N) = T$}. 
    \end{split}
\end{equation}
\noindent where $N$ is the white noise. More details are shown in Algorithm~\ref{alg:taser-opa}.
	
\begin{algorithm}
    \footnotesize
    \renewcommand{\algorithmicrequire}{\textbf{Input:}}
    \renewcommand{\algorithmicensure}{\textbf{Output:}}
	\caption{ALIF-OTA}~\label{alg:gen}
	\label{alg:taser-opa} 
	\begin{algorithmic}[1]
		\REQUIRE Target command $T$; target API SR; encoder $E$ and decoder $D$ of Tacotron2; vocoder V; threshold $Thr$; the number of particles $k$; the number of iterations $EpochMax$; the amplitude of white noise $\eta$ and the blurring parameters $\alpha$, $\beta$ and $\gamma$.
		\ENSURE $AttackExample$
        \STATE Initialize the value of $k$ particles $P=\{\delta_0, \delta_1, ..., \delta_k\}$ to $0$, and the velocity is random; 
        \STATE $Epoch=0; gloss=+\infty; ploss=\{+\infty_0, +\infty_1, ..., +\infty_k\}$
        \STATE $Emb = E(T)$
        \STATE $OriSpec = D(Emb)$
        \STATE $OriginalAudio = V(OriSpec)$
        \WHILE{$Epoch<EpochMax$}
            \FOR{$\delta_i \in P$}
                \STATE $AdvSpec_i = Blur(D(Emb + \delta_i), \alpha, \beta, \gamma$)
                \STATE $CandidateExample_i = V(AdvSpec_i)$
                \STATE Generate white noise $N$ with an amplitude of $\eta$
                \STATE $CandidateExample_i' = CandidateExample_i+N$
                \STATE $t_i = SR(CandidateExample_i)$
                \STATE $t_i' = SR(CandidateExample_i')$
                \IF{$t_i == T \And t_i' == T$}
                    \STATE $L_{Finali} = L_{Meli}+L_{Gatei}$
                \ELSE
                    \STATE $L_{Finali} = +\infty$
                \ENDIF
                \IF{$ploss_i>L_{Finali}$}
                    \STATE $ploss_i = L_{Finali}$
                    \STATE $pbest_i = \delta_i$
                \ENDIF
                \IF{$gloss>L_{Finali}$}
                    \STATE $gloss = L_{Finali}$
                    \STATE $gbest = \delta_i$
                    \STATE $AttackExample$ = $CandidateExample_i$
                \ENDIF
            \ENDFOR
            \STATE Update all particles according to $Thr$, $gbest$ and $pbest$ by Eq.\ref{eq:pso1} and \ref{eq:pso3}
            \STATE $Epoch++$
        \ENDWHILE
        \RETURN $AttackExample$
	\end{algorithmic} 
\end{algorithm}

\begin{table*}[htb]
\centering
\caption{Attack success rates of ALIF-OTL on commercial ASR APIs.}
\label{tab:API_Baseline}
\resizebox{0.9\textwidth}{!}{
\begin{threeparttable}
\begin{tabular}{@{}ccccccccccccccc@{}}
\toprule
\multicolumn{3}{c}{parameters}                  & \multicolumn{3}{c}{Amazon}       & \multicolumn{3}{c}{Azure}        & \multicolumn{3}{c}{iFLYTEK}      & \multicolumn{3}{c}{Tencent}      \\ \midrule
$\gamma$              & $\beta$               & $\alpha$ & online-SR & offline-SR & query & online-SR & offline-SR & query & online-SR & offline-SR & query & online-SR & offline-SR & query \\ \midrule
\multirow{2}{*}{0} & \multirow{2}{*}{0} & 0.25  & 12/12     & 6/12       & 32.8  & 9/12      & 3/12       & 34.8  & 10/12     & 4/12       & 33.8  & 12/12     & 5/12       & 33.6  \\
                   &                    & 0.3   & 12/12     & 5/12       & 30.8  & 10/12     & 5/12       & 33.6  & 12/12     & 4/12       & 33.9   & 12/12     & 4/12       & 33.1  \\ \midrule
\multirow{2}{*}{1} & \multirow{2}{*}{1} & 0.25  & 12/12     & 4/12       & 35.6  & 9/12      & 1/12       & 38.0  & 10/12     & 2/12       & 40.3  & 11/12     & 4/12       & 34.6  \\
                   &                    & 0.3   & 12/12     & 3/12       & 31.3   & 10/12     & 5/12       & 35.0  & 11/12     & 4/12       & 36.3  & 11/12     & 4/12       & 34.9  \\ \midrule
\multirow{2}{*}{1} & \multirow{2}{*}{2} & 0.25  & 11/12     & 4/12       & 35.6  & 10/12     & 4/12       & 36.8   & 10/12     & 1/12       & 39.2  & 12/12     & 6/12       & 33.4  \\
                   &                    & 0.3   & 12/12     & 5/12       & 32.4  & 9/12      & 3/12       & 35.2  & 10/12     & 2/12       & 36.4  & 12/12     & 5/12       & 34.3  \\ \midrule
2                  & 4                  & 1     & 12/12     & 4/12       & 30.4  & 11/12     & 5/12       & 36.6     & 11/12     & 2/12       & 34.6  & 12/12     & 7/12       & 29.8  \\ \midrule
\multicolumn{3}{c}{average}                     & 11.9/12     & 4.4/12       & 32.7   & 9.7/12      & 3.7/12       & 35.7  & 10.6/12     & 2.7/12       & 36.4   & 11.7/12     & 5.0/12          & 33.4  \\ \bottomrule
\end{tabular}
\begin{tablenotes}
    \footnotesize
    \item The scalar factor, represented as $\alpha$, is employed to attenuate the amplitude of frequencies within the initial 30 Mel filter banks. $\beta$ is the Mel filter bank index. We set the amplitude of frequencies within the first $\beta$ Mel filter banks to 0. $\gamma (>0)$ means the upper limit of the uniform noise amplitude we add to the Mel spectrogram. The terms "online-SR" and "offline-SR" represent the two variants of OTL attacks that require at most 50 and 1 query, respectively.
\end{tablenotes}
\end{threeparttable}
}

\end{table*}

\section{Experiments}
For ALIF-OTL, we first present a large-scale baseline evaluation on industry-grade cloud ASR APIs. Then, we pick the attack audio generated from the baseline study to launch the platform attack against the subtitling services. For ALIF-OTA, we attack the same online APIs in the over-the-line scenario but launch the attack in the physical environment. Additionally, we examine the feasibility of attacking VAs. Lastly, we conduct an impact factor study to investigate the overall performance of the ALIF-OTA attack.
\subsection{Experimental Setup}\label{sec:command}
\noindent \textbf{Target Commands and Text.} We aim to generate attack audio samples that can be successfully transcribed as different speech commands\footnote{\label{foot:commands}The commands include \emph{airplane mode on}, \emph{call 123}, \emph{cancel my alarm clock}, \emph{I need help}, \emph{navigate to my office}, \emph{send a message to my mom}, \emph{transfer the payment}, \emph{turn on the light}, \emph{unlock the door}, and \emph{what's the time}.} and common sentences\footnote{\label{foot:stentence}\emph{I can't take it anymore} and \emph{darn it}.}. We use the same target commands and text for both ALIF-OTL and ALIF-OTA evaluation. In this section, we mainly conduct evaluations based on a dataset consisting of ten commands and two sentences. To validate the efficacy of our method, we increase the dataset size for a more comprehensive evaluation, successfully generating attack audio samples using 32 commands and sentences in total. More details are provided in Appendix~\ref{sec:appendixa}.



\noindent \textbf{Parameter Settings.} The Mel spectrogram blurring and noise addition steps involve numerous hyperparameters: 1) $\gamma$ refers to the maximum value of the uniformly distributed noise. 2) $\beta$ is the Mel filter bank index. We set the amplitude of frequencies within the first $\beta$ Mel filter banks to 0. 3) $\alpha$ is the coefficient of energy attenuation in the low-frequency region of the Mel spectrogram. 4) $\eta$ indicates the magnitude of the white noise relative to the normalized synthesized audio waveform. 


\noindent \textbf{Targets.} 
\emph{OTL scenario.} We first conduct a baseline study: evaluating the performance of ALIF-OTL on commercial ASR APIs, including Amazon~\cite{AmazonTranscribe}, Microsoft Azure~\cite{AzureSTT}, iFLYTEK~\cite{iFLYTEKASR}, and Tencent~\cite{TencentASR}. Because Amazon, iFLYTEK~\cite{iFLYTEKASUBTITLE} and Microsoft~\cite{Clipchamp} also provide the subtitling service, we pick the adversarial audio samples in the baseline study to perform the platform attack. \emph{OTA scenario.} We launch the attack by playing the adversarial audio samples over the air in the environment. The audio samples are recorded and transcribed in three ways: 1) they are recorded by a common USB microphone connected to a laptop and are then transcribed by the four ASR APIs; 2) they are recorded and processed by a laptop running Microsoft's Cortana voice assistant; 3) Amazon Echo processes them. 


\noindent \textbf{Hardware.} In the OTA scenario, we use a Marshall EMBERTON II as the Bluetooth speaker for playing the adversarial audio. The laptop running Cortana is a Lenovo Thinkbook14+, and we use a Xiaomi phone connected to the Marshall speaker for playing audio. When attacking a standalone smart speaker, we use a third generation Amazon Echo Dot and play the attack audio via the Marshall speaker connected to the laptop. 







\noindent \textbf{Evaluation Metrics.}
We measure the attack effectiveness using the attack success rate (SR), the proportion of all attack audio successfully transcribed into the target command/text.  

\subsection{Evaluation of ALIF-OTL Attacks on Cloud Speech-to-Text APIs}\label{sec:evalalifotl}
\noindent \textbf{Baseline.} To evaluate the effectiveness of the ALIF-OTL attack, we perform the attack against four cloud ASR APIs. Each attack instance is generated after at most 50 queries to the target API. Specifically, for the \emph{online attack}, we only query the target ASR as long as the loss continues to decrease, and terminate after 50 iterations regardless. For the \emph{offline attack}, we only query the API once after 50 iterations. For both situations, the process also terminates if the perturbation magnitude exceeds the threshold. We generate ten audio instances for each command and select the best one. This number of instances is a parameter that can vary depending on the attack strategy. Table~\ref{tab:API_Baseline} shows the evaluation results. In the online attack, our method can generate attack examples for almost all target commands. The best attack success rate achieved was 95.8\% on the four APIs with only 35 queries per attack sample, using the parameter settings of $\alpha=1$, $\beta=4$, and $\gamma=2$. In offline attacks, our method reaches a success rate of 33.3\% for Amazon, Azure, and Tencent services with parameter settings of $\alpha = 0.3$, $\beta = 0$, and $\gamma = 0$. Though this seems low compared to the success rate of the online attacks, offline attacks do not require any API queries during the attack sample generation, which can limit the success rate due to the lack of feedback. The experimental results validate the effectiveness of the ALIF-OTL attack on commercial APIs. As a comparison, Devil's whisper~\cite{chen2020devilswhisper} failed to attack the Amazon API effectively, and its success rate was only 4/10, even with the confidence score. OCCAM~\cite{zheng2021black} can achieve a 100\% success rate on Azure, iFLYTEK, and Tencent but requires a significantly high number of queries (around 30,000). Our work achieves a success rate of 95.8\% with very few queries (about 35 for one instance), and the efficiency is 99.9\% and 97.7\% higher than that of OCCAM and Devil's whisper, respectively.


\begin{table}[t]
    \centering
    \caption{Success rates of ALIF-OTL attacks on subtitling services.}
    \label{tab:subtitle}   
    \resizebox{0.9\columnwidth}{!}{
    \begin{threeparttable}
\begin{tabular}{@{}cccccc@{}}
\toprule
\multicolumn{2}{c}{Amazon} & \multicolumn{2}{c}{Microsoft} & \multicolumn{2}{c}{iFLYTEK} \\ \midrule
digital-SR   & attack-SR   & digital-SR   & attack-SR  & digital-SR    & attack-SR   \\
12/12        & 5/12        & 10/12        & 5/10       & 11/12         & 5/11        \\ \bottomrule
\end{tabular}
 \begin{tablenotes}
    \footnotesize
    \item ``digital-SR" means the success rates of online attacks on the ASR APIs, which are the same as those in Table~\ref{tab:API_Baseline} ($\alpha=0.3, \beta=1, \gamma=1$). ``attack-SR" means the success rates of attacking the subtitling services.
    \end{tablenotes}
\end{threeparttable}
    }
\end{table}
\noindent \textbf{Platform Attacks.}
With the parameters where $\alpha = 0.3$, $\beta=1$, and $\gamma = 1$, we pick the attack samples that can successfully attack Amazon, Azure, and iFLYTEK, respectively, then use these adversarial audio samples to attack the online subtitling services. Windows provides the subtitling service via a software named Clipchamp, iFLYTEK only provides website entry without API, and we assume a close relationship between the Amazon ASR model and the subtitling model. Therefore, we perform a transfer attack rather than training attack audio samples tailored for each platform.   

The experimental results are shown in Table~\ref{tab:subtitle}. Almost half of the attack examples are still effective after transferring from speech recognition APIs to automatic subtitling APIs. The results verify the threat of our method to subtitling services. The performance can be further improved by training attack samples specifically targeting the Amazon subtitling API. 


\begin{table*}[ht]
    \centering
    \caption{Success rates of ALIF-OTA attacks on commercial ASR APIs.}
    \label{tab:ota}   
    \resizebox{0.7\textwidth}{!}{
    \begin{threeparttable}
\begin{tabular}{@{}cccccccccccc@{}}
\toprule
\multicolumn{4}{c}{parameters}                                               & \multicolumn{2}{c}{Amazon} & \multicolumn{2}{c}{Azure} & \multicolumn{2}{c}{iFLYTEK} & \multicolumn{2}{c}{Tencent} \\ \midrule
$\gamma$           & $\beta$           & $\alpha$          & $\eta$ & digital-SR   & record-SR   & digital-SR   & record-SR  & digital-SR    & record-SR   & digital-SR    & record-SR   \\ \midrule
\multirow{4}{*}{0} & \multirow{4}{*}{0} & \multirow{2}{*}{0.25} & 0.05       & 9/12         & 5/12        & 6/12         & 4/12       & 6/12          & 5/12        & 9/12          & 5/12        \\
                   &                    &                       & 0.1        & 11/12        & 7/12        & 6/12         & 6/12       & 4/12          & 3/12        & 10/12         & 5/12        \\
                   &                    & \multirow{2}{*}{0.3}  & 0.05       & 12/12        & 8/12        & 7/12         & 5/12       & 7/12          & 6/12        & 9/12          & 8/12        \\
                   &                    &                       & 0.1        & 12/12        & 7/12        & 5/12         & 5/12       & 6/12          & 5/12        & 11/12         & 6/12        \\ \midrule
\multirow{4}{*}{1} & \multirow{4}{*}{1} & \multirow{2}{*}{0.25} & 0.05       & 8/12         & 6/12        & 5/12         & 4/12       & 6/12          & 6/12        & 11/12         & 5/12        \\
                   &                    &                       & 0.1        & 8/12         & 6/12        & 5/12         & 4/12       & 5/12          & 5/12        & 9/12          & 7/12        \\
                   &                    & \multirow{2}{*}{0.3}  & 0.05       & 10/12        & 9/12        & 8/12         & 6/12       & 7/12          & 7/12        & 9/12          & 4/12        \\
                   &                    &                       & 0.1        & 9/12         & 8/12        & 7/12         & 7/12       & 3/12          & 3/12        & 9/12          & 5/12        \\ \midrule
\multirow{4}{*}{1} & \multirow{4}{*}{2} & \multirow{2}{*}{0.25} & 0.05       & 8/12         & 7/12        & 5/12         & 3/12       & 7/12          & 7/12        & 8/12          & 6/12        \\
                   &                    &                       & 0.1        & 8/12         & 7/12        & 6/12         & 6/12       & 5/12          & 5/12        & 7/12          & 5/12        \\
                   &                    & \multirow{2}{*}{0.3}  & 0.05       & 11/12        & 10/12       & 9/12         & 7/12       & 9/12          & 8/12        & 10/12         & 8/12        \\
                   &                    &                       & 0.1        & 12/12        & 9/12        & 8/12         & 8/12       & 5/12          & 4/12        & 9/12          & 6/12        \\ \midrule
\multirow{2}{*}{2} & \multirow{2}{*}{4} & \multirow{2}{*}{1}    & 0.05       & 12/12        & 12/12       & 10/12        & 8/12       & 11/12         & 9/12        & 10/12         & 8/12        \\
                   &                    &                       & 0.1        & 12/12        & 12/12       & 9/12         & 7/12       & 11/12         & 11/12       & 9/12          & 9/12        \\ \midrule
\multicolumn{4}{c}{average}                                                  & 10.1/12        & 8.1/12        & 6.9/12         & 5.7/12       & 6.6/12          & 6.0/12        & 9.3/12          & 6.2/12        \\ \bottomrule
\end{tabular}
 \begin{tablenotes}
    \footnotesize
    \item[1] $\eta$ is the amplitude of white noise we add to our attack examples.
    \item[2] ``digital-SR" refers to the success rates of ALIF-OTA attacks towards the APIs in the digital domain. ``record-SR" means the success rates of ALIF-OTA attacks in the physical world, where we play and record our attack examples and feed them to the APIs.
\end{tablenotes}
\end{threeparttable}
    }
\end{table*}

\begin{table}[ht]
    \centering
    \caption{Success rates of ALIF-OTA attacks on voice assistants.}
    \label{tab:ota2}   
    \resizebox{0.9\columnwidth}{!}{
\begin{threeparttable}
\begin{tabular}{@{}cccccccc@{}}
\toprule
\multicolumn{4}{c}{parameters}                                               & \multicolumn{2}{c}{Echo} & \multicolumn{2}{c}{Cortana} \\ \midrule
$\gamma$           & $\beta$           & $\alpha$          & $\eta$  & digital-SR & physical-SR & digital-SR   & physical-SR  \\ \midrule
\multirow{4}{*}{0} & \multirow{4}{*}{0} & \multirow{2}{*}{0.25} & 0.05       & 7/10       & 4/10        & 6/12         & 5/12         \\
                   &                    &                       & 0.1        & 9/10       & 4/10        & 6/12         & 6/12         \\
                   &                    & \multirow{2}{*}{0.3}  & 0.05       & 10/10      & 8/10        & 7/12         & 7/12         \\
                   &                    &                       & 0.1        & 10/10      & 6/10        & 5/12         & 4/12         \\ \midrule
\multirow{4}{*}{1} & \multirow{4}{*}{1} & \multirow{2}{*}{0.25} & 0.05       & 6/10       & 4/10        & 5/12         & 4/12         \\
                   &                    &                       & 0.1        & 6/10       & 3/10        & 5/12         & 4/12         \\
                   &                    & \multirow{2}{*}{0.3}  & 0.05       & 8/10       & 4/10        & 8/12         & 7/12         \\
                   &                    &                       & 0.1        & 7/10       & 6/10        & 7/12         & 7/12         \\ \midrule
\multirow{4}{*}{1} & \multirow{4}{*}{2} & \multirow{2}{*}{0.25} & 0.05       & 6/10       & 4/10        & 5/12         & 4/12         \\
                   &                    &                       & 0.1        & 7/10       & 4/10        & 6/12         & 5/12         \\
                   &                    & \multirow{2}{*}{0.3}  & 0.05       & 9/10       & 5/10        & 9/12         & 4/12         \\
                   &                    &                       & 0.1        & 10/10      & 5/10        & 8/12         & 7/12         \\ \midrule
\multirow{2}{*}{2} & \multirow{2}{*}{4} & \multirow{2}{*}{1}    & 0.05       & 10/10      & 7/10        & 10/12        & 4/12         \\
                   &                    &                       & 0.1        & 10/10      & 8/10        & 9/12         & 7/12         \\ \bottomrule
\end{tabular}
 \begin{tablenotes}
    \footnotesize
    \item ``digital-SR" refers to the success rates of ALIF-OTA attacks towards the APIs in the digital domain. ``physical-SR" means the success rates of attacking the VAs in the physical world.
\end{tablenotes}
\end{threeparttable}
    }
\end{table}

\subsection{Evaluation of ALIF-OTA Attacks on APIs and Voice Assistants.}
\noindent\textbf{Attacks on cloud ASR APIs.} We attack the same APIs of ALIF-OTL but launch the attack by playing attack audio samples out loud, using a speaker. We conduct the evaluation in a normal bedroom using a Marshall EMBERTON II speaker and use a fixed speaker-to-microphone distance of 15cm (the same setting as Zheng~et~al.~\cite{zheng2021black}). Table~\ref{tab:ota} demonstrates the attack performance of ALIF-OTA. The column of ``digital SR" refers to the success rate of our ALIF-OTA algorithm in the digital domain. In most parameter settings, our method can successfully generate almost all target commands. We play and record all the attack audio samples three times and provide them to the corresponding API. We consider an attack sample successful if at least one of the three recordings is transcribed by the targeted API to the intended target command. The results show that despite enduring interference in the physical domain, our attack can achieve a success rate of more than 50\% towards all the APIs under most parameter settings. Specifically, the success rate is 81.2\% under $\alpha=1$, $\beta=4$, $\gamma=2$ and $\eta=0.1$.

\noindent \textbf{Attacks on voice assistants.} Assuming the ASR systems behind the online API and the VA of the same company are similar, we use the adversarial audio samples generated from the API attack scenario to attack the Amazon Echo Dot and Microsoft Cortana (i.e., a transfer attack). Since Amazon Echo can't respond to ``Darn it!" and ``I can't take it anymore!", we generate one attack example for each of the remaining ten commands. Under the same environment settings as the cloud ASR API attack, we regard the attack as successful if the targeted command can be correctly transcribed within ten attempts (i.e., play each attack sample ten times). The results are in Table~\ref{tab:ota2}. Under different parameter settings, our attack can achieve an average success rate of up to 69.2\% on VAs (80\% on Echo and 58.3\% on Cortana), which illustrates the effectiveness of our OTA attack. Although Ni-OCCAM can also attack the VAs with a low cost, their success rates (average 50\% on Echo and Cortana) are lower.

\begin{table*}[htb]
    \centering
    \caption{Attack success rates of ALIF-OTA under different environments and device settings.}
    \label{tab:impact}   
    \resizebox{0.7\textwidth}{!}{
\begin{tabular}{@{}cccccccccccccc@{}}
\toprule
                               & speaker         & \multicolumn{4}{c}{EDIFIER}                            & \multicolumn{4}{c}{marshall}                                   & \multicolumn{4}{c}{JBL}                                \\ \cmidrule(l){2-14} 
                               & voice assistant & \multicolumn{2}{c}{Echo} & \multicolumn{2}{c}{Cortana} & \multicolumn{2}{c}{Echo} & \multicolumn{2}{c}{Cortana}         & \multicolumn{2}{c}{Echo} & \multicolumn{2}{c}{Cortana} \\
\multirow{-3}{*}{}             & distance         & SR         & dB          & SR          & dB            & SR         & dB          & SR  & dB                            & SR         & dB          & SR          & dB            \\ \midrule
                               & 15cm            & 4/7        & 76.04       & 5/7         & 76.54         & 6/7        & 76.1        & 7/7 & 74.59                         & 6/7        & 74.37       & 6/7         & 73.93         \\
                               & 30cm            & 2/7        & 70.19       & 6/7         & 69.13         & 4/7        & 69.56       & 6/7 & 67.56                         & 2/7        & 69.34       & 6/7         & 68.3          \\
                               & 50cm            & 3/7        & 64.56       & 7/7         & 64.8          & 3/7        & 64.13       & 6/7 & 62.94                         & 2/7        & 64.49       & 3/7         & 64.23         \\
                               & 100cm           & 3/7        & 60.57       & 7/7         & 60.9          & 2/7        & 59.34       & 5/7 & 59.09 & 3/7        & 60.19       & 5/7         & 60.76         \\
\multirow{-5}{*}{bedroom}      & 200cm           & 3/7        & 55.43       & 3/7         & 56.59         & 1/7        & 56.6        & 4/7 &                                    56.86 & 4/7        & 57.17       & 3/7         & 57.54         \\ \midrule
                               & 15cm            & 5/7        & 74.84       & 5/7         & 75.91         & 5/7        & 74.54       & 6/7 & 74.44                         & 3/7        & 72.03       & 3/7         & 71.06         \\
                               & 30cm            & 6/7        & 70.59       & 5/7         & 69.67         & 6/7        & 68.07       & 6/7 & 68.87                         & 1/7        & 67.71       & 2/7         & 66.53         \\
                               & 50cm            & 5/7        & 65.44       & 6/7         & 65.64         & 3/7        & 63.47       & 6/7 & 63.91                         & 1/7        & 64.17       & 3/7         & 63.54         \\
                               & 100cm           & 2/7        & 61.36       & 2/7         & 59.91         & 2/7        & 57.69       & 2/7 & 59.46                         & 1/7        & 60.43       & 2/7         & 59.76         \\
\multirow{-5}{*}{meeting room} & 200cm           & 2/7        & 56.39       & 3/7         & 56.95         & 0          & 56.69       & 0   & 54.96                         & 1/7        & 55.77       & 1/7         & 55.07         \\ \bottomrule
\end{tabular}
    }
\end{table*}

\begin{table}[htb]
    \centering
    \caption{Impact of ambient noise.}
    \label{tab:ambientnoise}   
    \resizebox{0.9\columnwidth}{!}{
        \begin{tabular}{@{}cccccc@{}}
        \toprule
            Voice assistant & $SNR\approx37$ & $SNR\approx25$ & $SNR\approx13$ & $SNR\approx3$\\ \midrule
            Echo & 6/7  & 6/7  & 3/7 & 3/7  \\
            Cortana & 7/7  & 5/7  & 3/7 & 2/7  \\ \bottomrule
        \end{tabular}
    }
\end{table}



\subsection{Impact of Various Factors on ALIF-OTA}

\noindent \textbf{Speaker dependency.} Different speakers have various hardware properties that affect the audio attributions. We intend to understand the sensitivity of audio samples generated by ALIF-OTL to the particular sound profiles of individual playback devices. Table~\ref{tab:impact} describes the attack success rate of our attack commands in different hardware and environmental settings. We pick all examples generated from ALIF-OTA using the parameters $\gamma=1$, $\beta=1$, $\alpha=0.3$, and $\eta = 0.1$ and test their performance using products from three well-known speaker manufacturers. The results demonstrate that our attack examples exhibit similar performances when played by the three speakers. All speakers achieve a success rate of higher than 57.1\% (4/7) on Echo and Cortana at a distance of 15cm. The only exception is using the JBL Pulse5 in the meeting room, which results in a success rate of less than 50\% (3/7) for Echo and Cortana.


\noindent \textbf{Evaluation in different rooms.} Different room types affect the propagation of acoustic signals differently. We pick two typical room types with different reverberation\footnote{A $5m*3.8m$ bedroom and a $5.5m*5.5m$ meeting room.} effects to evaluate their impact on our attack. The results show that our samples can get similar success rates within 50 centimeters in both rooms. When the distance increases to 1m and beyond, the performance of our commands becomes less effective in the meeting room than in the bedroom. We suspect this is due to the stronger reverberation in the meeting room since our samples were not explicitly processed for reverberation effect.

\noindent \textbf{Evaluation of different attack distances.} To evaluate the attack robustness with respect to propagation distance, we conduct the OTA attack across varying distances. As shown in Table~\ref{tab:impact}, when the distance is 15cm in the bedroom, the Marshall speaker achieves the best success rate, which is close to 100\% (6/7) for Echo and 100\% (7/7) for Cortana. The value drops to 42.9\% (3/7) and 85.7\% (6/7) at 50cm. When the distance reaches 2m, the success rates are 14.3\% (1/7) and 54.1\% (4/7). Despite the performance degradation, ALIF-OTA still achieves a success rate of higher than 50\% in the challenging 2m distance. To the best of our knowledge, this is the longest attack distance in studies on black-box audio attacks (same as Devil's Whisper). NI-OCCAM~\cite{zheng2021black}, which has a similar scenario to ours, achieved a success rate of 60\% on Cortana at a distance of 15cm and was not evaluated for performance at a longer distance. 

\noindent \textbf{Evaluation of ambient noise.} We evaluate the robustness of our attack against ambient noise by launching our examples across various levels of white noise. As Table~\ref{tab:ambientnoise} illustrates, our examples demonstrate strong performance at a signal-to-noise ratio (SNR) of roughly 20dB and above, similar to soft speech, but performance diminishes with an increase in ambient noise. This aligns with the SNR level reported in Devil's Whisper~\cite{chen2020devilswhisper} and When Evil Calls~\cite{Liu2022WhenEvilCalls}. Notably, even with a decrease in SNR to 3 dB, we can still achieve a success rate of more than 40\% when attacking Echo. This emphasizes the robustness of our attack.



\begin{table}[htb]
    \centering
    \caption{User study on audio comprehensibility.}
    \label{tab:userstudy1}   
    \resizebox{0.9\columnwidth}{!}{
        \begin{threeparttable}
        \begin{tabular}{@{}cccccc@{}}
        \toprule
                & Non-native Speakers & Native Speakers & All volunteers\\ \midrule
            Score & 0.77  & 1.70  & 1.0   \\
            CER & 0.79  & 0.45  & 0.71   \\
            WER & 0.88  & 0.55  & 0.80  \\ \bottomrule
        \end{tabular}
         \begin{tablenotes}
    \footnotesize
    \item ``Score" is the average score calculated based on feedback provided by all volunteers. In addition to score the intelligibility, participants are asked to transcribe audio samples. ``CER" and ``WER" are character error rate and word error rate obtained by calculating the difference between their transcription and the ground truth text.
    \end{tablenotes}
\end{threeparttable}
    }
\end{table}

\begin{table}[t]
    \centering
    \caption{User study on blurring ablation.}
    \label{tab:userstudy2}   
    \resizebox{0.9\columnwidth}{!}{
    \begin{threeparttable}
        \begin{tabular}{@{}cccccc@{}}
        \toprule
            & L.P. Only & Beta Only & Gamma Only & Alpha Only & All Mel Blurring\\ \midrule
            Non-native Speaker & 1.84 & 3.68 & 3.68 & 2.93 & 2.96 \\
            Native Speaker & 2.84 & 3.96 & 3.76 & 3.04 & 3.24 \\
            All volunteers & 2.09 & 3.75 & 3.7 & 2.96 & 3.03  \\\bottomrule
        \end{tabular}
    \begin{tablenotes}
    \footnotesize
    \item Each column depicts the mean intelligibility score of audio samples produced by removing specific components from the ALIF pipeline. This illustrates the influence of our method's various components on user comprehension. ``L.P. Only" indicates audio samples with solely linguistic feature perturbation. ``Beta Only" means we only eliminate extremely low frequencies during the generation of attack audio samples. ``Gamma Only" and ``Alpha Only" follow a similar pattern. ``All Mel Blurring" signifies the application of all Mel spectrogram blurring techniques. It should be noted that linguistic feature perturbation is exclusively implemented in the ``L.P. Only" category.
    \end{tablenotes}
\end{threeparttable}
    }
\end{table}




\subsection{User Study}
We carried out a user study to gain deeper insight into human perception of the attack examples we generated. The study comprises two sections: audio incomprehensibility and ablation of different ALIF components. We recruited a varied group of 20 volunteers, aged 22 to 32, all with normal hearing abilities. Notably, five participants are native English speakers.

\noindent \textbf{Audio comprehensibility.} We selected 32 distinct command examples\footnote{ The audio samples include all commands and sentences in Footnote~\ref{foot:commands}, \ref{foot:stentence} and Appendix~\ref{sec:appendixa}.} produced by ALIF-OTL under the parameter settings of $\alpha=0.3$, $\beta=1$, and $\gamma=1$. We recruited volunteers to assess each sample using a 0-4 intelligibility scale\footnote{Scores range from 0 to 4; 0: completely incomprehensible, 1: few parts are understandable, 2: some parts are understandable, 3: most parts are understandable, and 4: completely understandable.} and subsequently transcribe them. The results are presented in Table~\ref{tab:userstudy1}. On average, participants assigned an intelligibility score of 1.0. Their transcriptions achieved a CER of 71\% and a WER of 80\%, indicating a significant deviation from the original commands. These statistics confirm the poor intelligibility of the audio samples.




\noindent \textbf{Distortion effect of ALIF's each component.} We aim to determine various components' effects on audio comprehensibility: perturbations on linguistic features and three Mel spectrogram blurring components. To do this, we selected five examples and ablated each component, resulting in 25 audio samples. We then asked volunteers to rate the comprehensibility of these audios and collected their feedback. As indicated in Table~\ref{tab:userstudy2}, the perturbation on linguistic features, which received the lowest score of 2.09, was identified as the most critical factor impacting human perception.

We conducted the ablation study to show the effects of the different components on user comprehension. Our ALIF algorithm ensures all generated attack samples can be correctly transcribed to target commands. Therefore, this experiment also shows that ablating various parts of the attack pipeline does not have a significantly negative effect on attack success.

\subsection{Long-Term Effectiveness of ALIF}
We validate the robustness of ALIF attacks against model updates by evaluating the long-term effectiveness of our attack audio samples. 

We generate 50 to 100 effective attack examples targeting each of the four commercial APIs on October $1^{st}$, 2022, and test them after three months. The success rates of our attack audios on different dates are presented in Table~\ref{tab:time}. Although the success rate on iFLYTEK decreases by about 25\%, the success rates on Amazon and Azure drop by less than 10\%. Notably, the attack performance on Amazon remains unchanged after three months. Furthermore, after extending the testing period by an additional half month, the performance across all ASRs remained steady, with multiple examples of each command successfully executing the attack. 

ALIF was employed to optimize the added perturbation to the linguistic feature to decrease audio comprehensibility. This is in contrast to existing black-box methods that add perturbation to shift the command signal towards another unsuspicious audio sample, bringing the signal closer to the ASR's decision boundary. Although ASR updates may still affect ALIF's performance, it has shown relative resilience.

\begin{table}[htb]
    \centering
    \caption{Change of the success rates of ALIF-OTL over time.}
    \label{tab:time}   
    \resizebox{0.9\columnwidth}{!}{
        \begin{tabular}{@{}cccccc@{}}
        \toprule
             API  & Amazon & Azure & iFLYTEK & Tencent \\ \midrule
            2022.10.01  & 100\% & 100\% & 100\% & 100\% \\ 
            2023.01.01  & 100\% & 91.53\% & 74.03\% & 94.12\% \\ 
            2023.01.06  & 100\% & 91.53\% & 74.03\% & 94.12\% \\ 
            2023.01.11  & 100\% & 91.53\% & 74.03\% & 94.12\% \\ 
            2023.01.16  & 100\% & 91.53\% & 74.03\% & 94.12\% \\ \bottomrule
        \end{tabular}
    }
\end{table}



\subsection{Comparisons with related adversarial attacks}
The key distinction between our work and existing AE studies is the way of constructing adversarial audio samples. Our work functions uniquely compared to existing works to improve the query efficiency and robustness against model updates. Existing works add noise to the raw waveform to mislead the ASR. Their training objective is to ensure the ASR can be misled while minimizing the waveform-level perturbation, as formulated in Equation~\ref{eq:blackboxAE}. In contrast, our approach generates audio signals that are incomprehensible to humans but recognizable by ASR. The training objective is to ensure that the ASR can recognize the command while maximizing the perturbations.

We compare our work and existing black-box AE attacks, as demonstrated in Table~\ref{tab:comparisonswithSOTA}. Because the objectives of adding perturbation differ, we do not compare aspects such as the perturbation level. Our approach shows an evident advantage in query efficiency and attack distance.

\begin{table}[t]
    \centering
    \caption{Comparisons with related adversarial audio attacks.}
    \label{tab:comparisonswithSOTA}   
    \resizebox{0.9\columnwidth}{!}{
    \begin{threeparttable}
\begin{tabular}{@{}ccccccc@{}}
\toprule
Attacks         & Gradient & Conf & Targeted & Over the air & Distance & Query \\ \midrule
CommanderSong~\cite{yuan2018commandersong}  & $\checkmark$      &                  & $\checkmark$      &       &       & 1000  \\ \midrule
Hidden voice~\cite{Carlini2016Hidden}    &          &                  & $\checkmark$      & $\checkmark$     &  300 cm    & 5000  \\ \midrule
Devil's whisper~\cite{chen2020devilswhisper} &          & $\checkmark$              & $\checkmark$      & $\checkmark$   & 5-200 cm    & 1500  \\ \midrule
OCCAM~\cite{zheng2021black}           &          &                  & $\checkmark$      &      &        & 30000 \\ \midrule
TAINT~\cite{Liu2022WhenEvilCalls}           &          &                  & $\checkmark$      & $\checkmark$     & 50 cm (VoIP)     & 1500  \\ \midrule
ALIF-OTL       &          &                  & $\checkmark$      &        &      & \textbf{50}    \\ \midrule
ALIF-OTA       &          &                  & $\checkmark$      & $\checkmark$    & 15-200 cm     & \textbf{400}   \\ \bottomrule
\end{tabular}
 \begin{tablenotes}
    \footnotesize
    \item ``Gradient" / ``Conf" means whether the method needs the gradient/confidence score of the target model. ``Targeted" denotes whether the attack is targeted or non-targeted. ``Over the air" means whether the attack can work in the physical world. ``Distance" is the distance between speaker and microphone in experiments (\cite{Liu2022WhenEvilCalls} launches the attack through VoIP). ``Query" means the total number of queries for one attack example.
    \end{tablenotes}
\end{threeparttable}
    }
\end{table}

\section{Related Work}
In this section, we survey studies related to speech command injection attacks. Note that we mainly cover research focusing on signal generation and do not include the work utilizing hardware property~\cite{zhang2017dolphinattack, Roy2017Backdoor,Roy2018NDSILongrange}.

\subsection{White-Box Adversarial Example}
In recent years, extensive studies apply AE techniques to the speech domain to achieve command injection. Carlini~et~al.~\cite{carlini2018AEs} first successfully generated the attack examples towards Deepspeech, an open-sourced end-to-end ASR platform. CommanderSong~\cite{yuan2018commandersong} tried to embed the malicious voice command into songs so that it won’t attract human attention but could mislead Kaldi successfully. However, these works are fragile in the physical world. Specifically, the adversarial examples will fail when being played over the air. To address this problem, Yakura~et~al. and Imperio~\cite{Yakura2018Adversarial, Schonherr2019imperio} simulated the transformations caused by the physical world and introduced these transformations into the process of adversarial example generation. Metamorph~\cite{Chen2020Metamorph} revealed that the signal distortion in the physical world is mainly caused by the device and channel frequency selectivity. They proposed a “generate-and-clean” two-phase design. To mitigate the human perception of adversarial perturbation, Schönherr~et~al.~\cite{Schonherr2019} generated adversarial examples based on psychoacoustic hiding. Inspired by the universal adversarial perturbations in the image domain~\cite{Moosavi2016Universal}, Neekhara~et~al.~\cite{Neekhara2019Universal} generated the universal perturbations in the speech domain and successfully attacked Mozilla DeepSpeech. AdvPulse~\cite{liAdvPulse2020} also proposed universal, synchronization-free adversarial perturbations which make the attack scenario more practical.



\subsection{Black-Box Adversarial Example}
Despite the success of adversarial examples in speech recognition attacks, the white-box premise of many tools poses a significant practicality concern; knowledge of the architecture and parameters of the model is unrealistic for a real-world attack scenario. Therefore, recently, black-box attacks have become an active research area. Taori~et~al.~\cite{Taori2018Adversarial} combined the approaches of genetic algorithms with gradient estimation to attack Deepspeech. This approach has a low success rate and is ineffective on commercial models. Devil’s Whisper~\cite{chen2020devilswhisper} utilized the confidence scores of the commercial ASR APIs. They built a substitute model to approximate the target model and launch the attack. However, most commercial APIs only return the final results without any confidence scores, which limits the practicality of this method. OCCAM~\cite{zheng2021black} is the first approach that attacks commercial ASR APIs successfully in the completely black-box scenario, i.e., no confidence scores are required. However, the cost of this method is high. A recent work, TAINT~\cite{Liu2022WhenEvilCalls}, considers the impact of the VoIP channel and uses gradient estimation to generate adversarial examples, resulting in a more robust and efficient attack.
\subsection{Hidden Voice Attacks}
Besides adversarial examples, “hidden voice command” is another line of attack method targeting speech recognition systems. Cocaine Noodles~\cite{vaidya2015cocaine} first used the inverse MFCC technique to create the attack sample which will be recognized by the devices but won’t be understood by humans. Carlini~et~al.~\cite{Carlini2016Hidden} extended it to the white-box scenario and proposed a “hidden voice command” attack that can’t be understood at all. Abdullah~et~al.~\cite{Abdullah2019Practical} then exploited signal processing algorithms to make hidden voice commands more practical. 


\section{Discussion}
\subsection{Defense against ALIF.}
\noindent\textbf{Downsampling.} Downsampling is a commonly-used method to neutralize AE attacks. According to Nyquist's theorem, the high-frequency component of the original audio will be removed after downsampling. Since AE adds perturbation at the high-frequency area, downsampling can effectively defend against existing black-box attacks. Liu~et~al.~\cite{Liu2022WhenEvilCalls} downsampled attack audio to 16 kHz, then upsampled to 48 kHz, causing none of their attacks to succeed. When OCCAM~\cite{zheng2021black} downsampled signals to 12 kHz and then upsampled back to 16 kHz, their attack fails, and the success rate of NI-OCCAM decreases to 30\%. To evaluate the robustness of ALIF audio against downsampling, we pick about 10 ALIF audio samples (one sample for each speech command) and perform the same 12 kHz downsampling action, resulting in a success rate of 38.64\%. To further push the limit, we perform downsampling to 8 kHz and still achieve a success rate of 20.45\%. The experimental results are shown in Table~\ref{tab:downsampling}, which demonstrate that ALIF has better robustness against downsampling, which makes sense because ALIF mainly affects the frequency range related to speech, which is usually below 8 kHz and has less dependence on higher frequencies. Overall, downsampling is an effective defense. ALIF could be more robust if the attacker knows the downsampling strategy and adapts the audio generation, but this is a strong assumption.



\begin{table}[t]
\centering
\caption{Success rates after downsampling the examples. }
\label{tab:downsampling}   
\resizebox{0.9\columnwidth}{!}{
\begin{threeparttable}
\begin{tabular}{cccccc}
\toprule
API   & Amazon & Azure & iFLYTEK & Tencent & total   \\ \midrule
8kHz  & 3/12   & 3/10  & 1/11    & 2/11    & 20.45\% \\
12kHz & 1/12   & 4/10  & 7/11    & 5/11    & 38.64\% \\ \bottomrule
\end{tabular}

\begin{tablenotes}
    \footnotesize
    \item We use the attack audio samples generated to evaluate ALIF-OTL performance in Section~\ref{sec:evalalifotl}. With the parameters set at $\alpha=0.3$, $\beta=1$, and $\gamma=1$, 12, 10, 11, and 11 audio samples were successfully generated to attack Amazon, Azure, iFLYTEK, and Tencent ASR, respectively.
\end{tablenotes}
\end{threeparttable}
}
\end{table}

\begin{table}[t]
\centering
\caption{Success rates after filtering the examples.}
\label{tab:filtering}   
\resizebox{0.9\columnwidth}{!}{
\begin{tabular}{cccccc}
\toprule
API   & Amazon & Azure & iFLYTEK & Tencent & total   \\ \midrule
High-Pass 500Hz  & 1/12   & 3/10  & 0/11    & 2/11    & 13.64\% \\
Low-Pass 4000Hz  & 1/12   & 5/10  & 3/11    & 3/11    & 27.27\% \\
Low-Pass 6000Hz & 2/12   & 5/10  & 5/11    & 4/11    & 36.36\% \\ \bottomrule
\end{tabular}
}
\end{table}



\noindent \textbf{Frequency filtering.} Apart from downsampling, we evaluate the impact of filtering on mitigating our attacks. We test three potential frequency filters: a 4000Hz low-pass filter, a 6000Hz low-pass filter, and a 500 Hz high-pass filter. Table~\ref{tab:filtering} displays that the three methods successfully defended against our attack. The 500Hz high-pass filter performed best, with only 1, 3, 0, and 2 attack commands accurately recognized by the respective ASRs. 

\noindent \textbf{Adversarial training.} 
In addition to downsampling, the defender could utilize adversarial training to enhance model robustness~\cite{papernot2016Distillation,papernot2017practical}. Because ALIF is based on a different working mechanism than AE attacks, hypothetically, the effect of typical adversarial training on the input waveform domain is limited. However, embedding-level adversarial training is potentially effective. The models are not accessible to perform adversarial training because we primarily attack commercial ASRs in this paper. We leave studies on this new type of adversarial training for future work. 

\begin{figure}[t]
    \begin{minipage}[t]{\linewidth}
        \centering          
        \includegraphics[width=0.9\columnwidth]{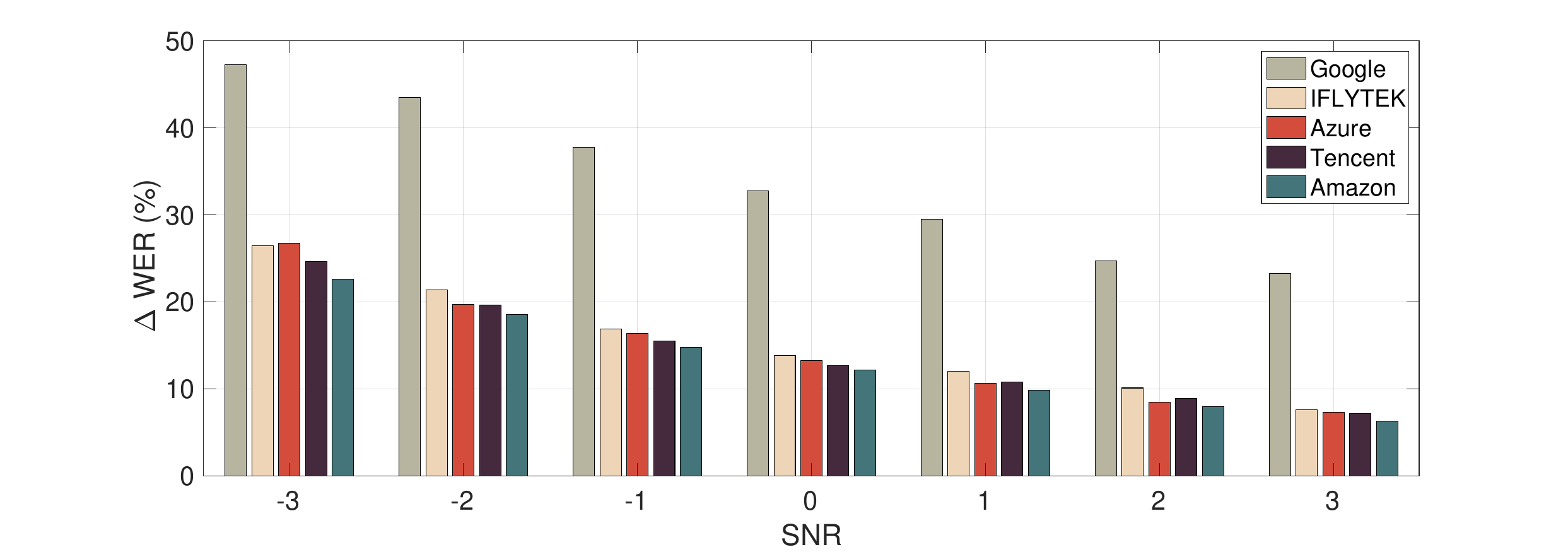} 
        \caption{Robustness of commercial ASRs against white noise}
        \label{fig:robustness_against_white_nosie}
    \end{minipage}
    \begin{minipage}[t]{\linewidth}
        \centering      
        \includegraphics[width=0.9\columnwidth]{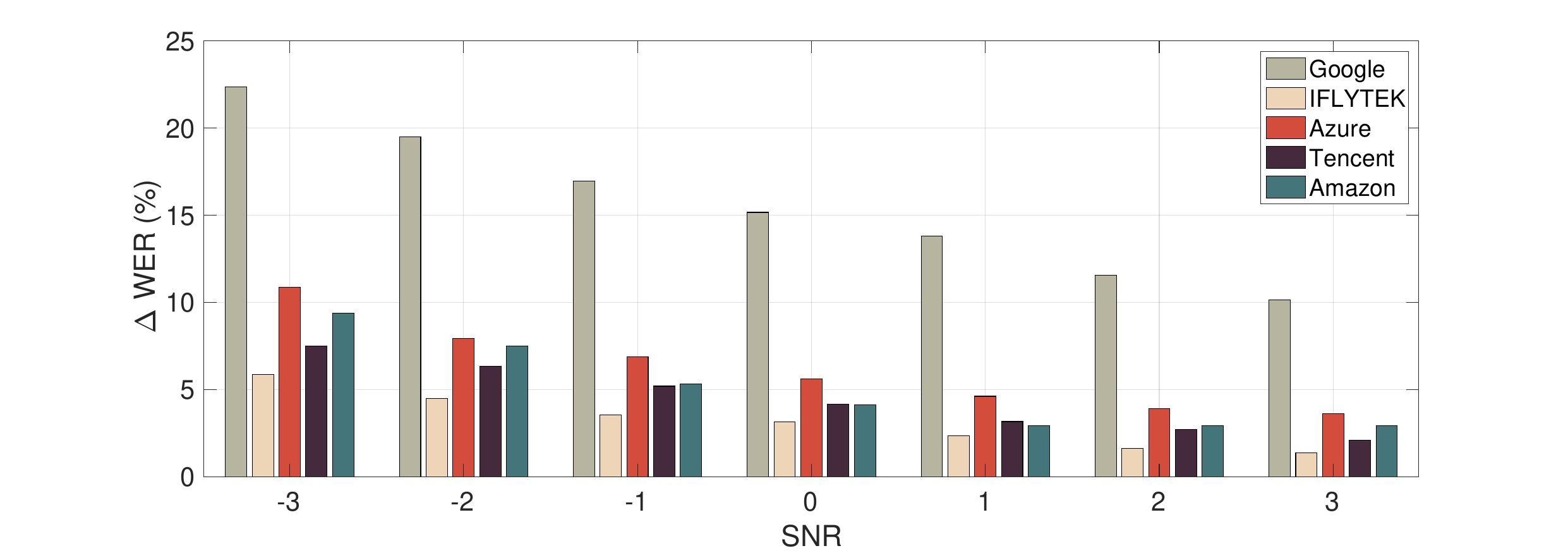}  
        \caption{Robustness of commercial ASRs against event noise}
        \label{fig:robustness_against_event_nosie}
    \end{minipage}
\end{figure}



\subsection{Attack Success and ASR Robustness}



Black-box attack training follows a specific paradigm in which a target phase is perturbed to both make the audio sound different as well as to limit a victim's awareness of the attack. The limit on the perturbation energy is no longer primarily designed for imperceptibility as in the white-box attack. Instead, it aims to prevent the transcription from being different than the target phrase. In this context, it is possible that the success of a black-box attack also depends on the ASR's robustness to noise. 

To investigate if a correlation exists, we test the robustness of five commercial ASRs and study the connection between the attack performance of ALIF-OTL and the noise robustness. We randomly choose 200 sentences from the LibriSpeech test-clean subset and mix them with Gaussian white noise and event noise under different SNRs, respectively. Then, we evaluate them using ASRs and calculate the increase of word error rate ($\Delta$WER) caused by white noise, where smaller $\Delta$WER indicates better robustness, and vice versa. Figure~\ref{fig:robustness_against_white_nosie} and Figure~\ref{fig:robustness_against_event_nosie} demonstrate that Amazon is fairly robust while Google is the most vulnerable one, which is consistent with some of our observations during the attack sample generation. That is, we can generate attack samples against Amazon API relatively easily but cannot generate attack examples of high quality against Google API. The experimental results suggest the possibility of \emph{the more robust the ASR is, the more vulnerable it is to our attack.} We will conduct a more comprehensive study to verify the assumption in the future.

\noindent \subsection{Limitations}
In this paper, we utilize PSO as the primary optimization method for ALIF-OTA and have not explored other techniques. Different optimization techniques can result in various search routes in the problem space, and it is worth more investigation.

In both algorithms for ALIF-OTL and ALIF-OTA, we use a fixed threshold to control the amplitude of perturbation added to the linguistic feature, which is not precise enough, since an ASR service usually recognizes different commands with various accuracy, indicating ASRs have different sensitivity to command contents. Setting the threshold dynamically depending on the attack command is more practical. We leave it for future work.




\section{Conclusion}
In this paper, we propose the first black-box adversarial audio attack methods that are highly efficient and robust against model updates: ALIF-OTL and ALIF-OTA. The two attack schemes are based on ALIF, a novel adversarial linguistic feature-based attack pipeline. We directly add perturbation to the low dimensional manifold where the decision boundary lies to generate adversarial audio samples, which overcomes the inherent shortcomings of conventional black-box AEs: query inefficiency and the vulnerability to model updates. ALIF-OTL only requires an average of 35 queries to generate attack audios against cloud ASR APIs. Experiments show ALIF-OTL is effective against four well-known commercial cloud ASRs, achieving an attack success rate of 95.8\%.  ALIF-OTA uses the PSO method to incorporate environmental interference into the training and only requires 400 queries for attack sample generation. Experiments show the efficacy of our attack audio samples in attacking four commercial ASRs and two VAs in the physical playback environment, achieving average success rates of 81.3\% and 69.2\%, respectively. Critically, the test-of-time experiment verifies the long-term effectiveness of the ALIF attack, indicating its robustness to model changes.

\section*{Acknowledgment}
This work is partially supported by the National Key R\&D Program of China (Grant No. 2020AAA0107700), the National Natural Science Foundation of China (Grant No. 62172359, 61972348, and 62102354), the National Key R\&D Program of China (Grant No. 2021ZD0112803), the Funding for Postdoctoral Scientific Research Projects in Zhejiang Province (Grant No. ZJ2021139), the Fundamental Research Funds for the Central Universities (Grant No. 2021FZZX001-27), and the Hangzhou Leading Innovation and Entrepreneurship Team (TD2020003).

\ifCLASSOPTIONcaptionsoff
  \newpage
\fi



\bibliographystyle{IEEEtran}
\bibliography{reference}
\clearpage

\appendices

\section{Additional Evaluations}\label{sec:appendixa}
We expand the command dataset size to demonstrate our method's efficacy. Besides the ten commands and two sentences discussed in the main paper, we also produce adversarial audio samples employing an additional 20 commands. The results are shown in Table~\ref{tab:moredata}.
\makeatletter
\setlength{\@fptop}{5pt}
\makeatother
\begin{table*}[!t]
    \centering
    \caption{Evaluation of ALIF on a larger dataset.}
    \label{tab:moredata}   
    \resizebox{0.9\textwidth}{!}{
    \begin{threeparttable}
        \begin{tabular}{@{}ccccccccccc@{}}
        \toprule
           & Echo & Cortana & Amazon-online & Amazon-offline & Azure-online & Azure-offline & iFLYTEK-online & iFLYTEK-offline & Tencent-online & Tencent-offline \\ \midrule
        Clear notification       & $\checkmark$  & -  & $\checkmark$  & -  & $\checkmark$  & -  & $\checkmark$  & -  & $\checkmark$  & - \\
        Good morning             & $\checkmark$  & $\checkmark$  & $\checkmark$  & -  & $\checkmark$  & -  & $\checkmark$  & -  & $\checkmark$  & - \\
        How old are you          & $\checkmark$  & $\checkmark$  & $\checkmark$  & -  & $\checkmark$  & -  & $\checkmark$  & -  & $\checkmark$  & $\checkmark$ \\
        Play music               & $\checkmark$  & $\checkmark$  & $\checkmark$  & -  & $\checkmark$  & -  & $\checkmark$  & -  & $\checkmark$  & - \\
        Sing me happy birthday   & $\checkmark$  & -  & $\checkmark$  & $\checkmark$  & $\checkmark$  & $\checkmark$  & $\checkmark$  & -  & $\checkmark$  & $\checkmark$ \\
        Take a picture           & $\checkmark$  & -  & $\checkmark$  & -  & $\checkmark$  & $\checkmark$  & $\checkmark$  & $\checkmark$  & $\checkmark$  & - \\
        Tell me a story          & $\checkmark$  & -  & $\checkmark$  & $\checkmark$  & $\checkmark$  & -  & $\checkmark$  & -  & $\checkmark$  & $\checkmark$ \\
        What's the weather       & $\checkmark$  & $\checkmark$  & $\checkmark$  & -  & $\checkmark$  & -  & $\checkmark$  & $\checkmark$  & $\checkmark$  & - \\
        Where is my car          & $\checkmark$  & -  & $\checkmark$  & -  & $\checkmark$  & -  & $\checkmark$  & -  & $\checkmark$  & $\checkmark$ \\
        Where is my home         & $\checkmark$  & $\checkmark$  & $\checkmark$  & -  & $\checkmark$  & -  & $\checkmark$  & $\checkmark$  & $\checkmark$  & $\checkmark$ \\
        Ask me a question        & $\checkmark$   & $\checkmark$   & $\checkmark$  & -  & $\checkmark$  & -  & $\checkmark$  & $\checkmark$  & $\checkmark$  & $\checkmark$ \\
        Clean my room            & $\checkmark$   & -   & $\checkmark$  & -  & $\checkmark$  & -  & $\checkmark$  & -  & $\checkmark$  & - \\
        Find a hotel             & -   & $\checkmark$   & $\checkmark$  & $\checkmark$  & $\checkmark$  & -  & $\checkmark$  & -  & $\checkmark$  & - \\
        Make it warmer           & $\checkmark$   & -   & $\checkmark$  & -  & $\checkmark$  & -  & -  & -  & $\checkmark$  & - \\
        Open the box             & $\checkmark$   & -   & $\checkmark$  & -  & $\checkmark$  & -  & $\checkmark$  & -  & $\checkmark$  & $\checkmark$ \\
        Open the website         & $\checkmark$   & $\checkmark$   & $\checkmark$  & $\checkmark$  & $\checkmark$  & -  & $\checkmark$  & -  & $\checkmark$  & - \\
        Reading a book           & $\checkmark$   & -   & $\checkmark$  & -  & $\checkmark$  & -  & $\checkmark$  & -  & $\checkmark$  & - \\
        Show me the money        & $\checkmark$   &  $\checkmark$  & $\checkmark$  & $\checkmark$  & $\checkmark$  & $\checkmark$  & $\checkmark$  & -  & $\checkmark$  & - \\
        Turn off the computer    & $\checkmark$   & $\checkmark$   & $\checkmark$  & $\checkmark$  & $\checkmark$  & -  & $\checkmark$  & -  & $\checkmark$  & - \\
        Turn on bluetooth        & $\checkmark$   & -  & $\checkmark$  & -  & $\checkmark$  & -  & $\checkmark$  & -  & $\checkmark$  & - \\ \midrule
        Total        & 19/20   & 10/20   & 20/20  & 6/20  & 20/20  & 3/20  & 19/20  & 4/20  & 20/20  & 7/20 \\ \bottomrule
        \end{tabular}
         \begin{tablenotes}
            \footnotesize
            \item The terms "Echo" and "Cortana" refer to OTA attacks on Echo and Cortana, respectively, which is the same as Table~\ref{tab:ota2}. The terms "xxx-online" and "xxx-offline" represent the two variants of OTL attacks executed against various APIs, aligning with the information in Table~\ref{tab:API_Baseline}. The checkmark symbol ``$\checkmark$" indicates the successful generation of examples and execution of attacks on the target API or VA. All examples adhere to a parameter setting of $\alpha=0.3, \beta=1, and \gamma=1$. In cases of OTA attacks, $\eta=0.1$ is also applied.
            \end{tablenotes}
    \end{threeparttable}
    }
\end{table*}


\newpage
\section{Meta-Review}

\subsection{Summary}
The authors propose two attack schemes: ALIF-OTL and ALIF-OTA to generate adversarial black-box attack pipelines based on the linguistic feature space of TTS. The authors propose over-the-line and over-the-air approaches and implement a comprehensive analysis against commercial ASRs, reaching more than 95\% query efficiency improvement compared with the state-of-the-art.

\subsection{Scientific Contributions}
\begin{itemize}
\item Provides a Valuable Step Forward in an Established Field
\end{itemize}

\subsection{Reasons for Acceptance}
\begin{enumerate}
\item The paper identifies and uses linguistic structures present in Text-to-Speech algorithms as a means to improve the generation of adversarial audio for their attacks. This allows for the model during the training process to target audio features that human listeners are sensitive to, thus allowing for the model to be trained faster and with less resources.
\item The work highlights an interesting possible external source of a priori knowledge for future work to explore, namely the use of linguistic structures found in TTS.
\end{enumerate}

\subsection{Noteworthy Concerns} 
\begin{enumerate} 
\item The authors only provide a small, informal user study consisting of 20 volunteers who evaluated 32 audio samples. Given the attacker's goal in this work is to create audio samples that are incomprehensible to human listeners, but still comprehensive to machines, this user study is insufficient. Additionally, several of the reviewers of this work found the audio samples easily intelligible. Thus, the overall incomprehensibility of the created audio samples cannot be determined at this time.
\item One of the main contributions of this work is the improved training efficiency of both techniques. Despite this, the paper uses a small number of audio samples, two sentences and ten commands\footnote{The authors added an additional 20 commands in the Appendix.}, to evaluate the work. While this number is inline with previous work, we would expect that with efficiency improvements we would see larger datasets used for evaluation.
\item The real-world attack scenario of this work is lacking. In Section 7 the authors evaluate the performance of the OTA attack and find that the attack success drops significantly past 15 cm (less than 50\% at 15 cm for certain ASRs). Additionally, on their website and in their rebuttal the authors detail the attack "performs well at a SNR of $\sim$20[dB]." In the context of acoustics, this is the difference between a bedroom and a restaurant\footnote{Britannica - https://www.britannica.com/science/sound-physics/The-decibel-scale}. This detail gives a possible explanation to the performance loss seen as the attack distance increased since acoustic energy reduces with the inverse square law. These aspects of the paper bring into question the deployability of the proposed attacks.
\item Additional concerns were raised over the selection and effects of the different parameters used in the attacks. An ablation study was provided, however, it focused on evaluated human comprehension, not attack success.
\end{enumerate}

\section{Response to the Meta-Review}
We are grateful to our reviewers and the S\&P 2024 program committee for accepting our paper, and we appreciate the effort put into summarizing and identifying the limitations of our work.

We would like to address some points raised in the meta-review:

\textbf{Comment 2} notes that the reviewers would like us to use larger command datasets for generating attack audio samples.

We have taken this into consideration and increased our dataset size to include an additional 20 commands, making a total of 32 commands and sentences. These commands were carefully selected from common user interactions, thereby representing significant security threats. The successful generation of attack audio samples from this corpus affirms the effectiveness of our method, and we believe that further increase would not yield additional technical challenges. Thus, we believe that the current size is adequate for the purpose of our study.

\textbf{Comment 3} points out the lack of real-world attack scenarios.

We would like to note that the Signal-to-Noise Ratio (SNR) and attack distance requirements of our study align with related work. We plan to enhance the practicality of our method in our future research.

\textbf{Comment 4} notes that the ablation study focuses on evaluating human comprehension rather than attack success.

Our attack algorithm can ensure successful attacks under various ablation conditions, making this aspect less crucial for evaluation in our study. Therefore, we did not ablate components of the attack pipeline to study the impact on attack success.

\end{document}